\documentclass[aps,prl,twocolumn,showpacs,amsmath,amssymb]{revtex4-1}

\usepackage{graphicx}
\usepackage{color}
\usepackage{epstopdf}
\usepackage{amsmath}
\allowdisplaybreaks[4]
\usepackage{bm}
\usepackage{hyperref}
\usepackage{cleveref}
\usepackage{tcolorbox}
\usepackage{braket}
\usepackage[export]{adjustbox}
\usepackage[hang]{subfigure}
\usepackage{MnSymbol}
\usepackage{makecell}
\usepackage{float}
\usepackage{booktabs}
\usepackage{upgreek}

\begin{document}
	%\title{Quasicrystals and excitations in dipolar Bose-Einstein condensates with optical feedback}

	%\title{Optically driven quasicrystals and excitations in dipolar Bose-Einstein condensates}
	
	%\title{Self-organized quasicrystals and excitations in dipolar Bose-Einstein condensates assisted by optical feedback}
	
	\title{Self-organized quasicrystals and their excitations in dipolar Bose-Einstein condensates via optical feedback}
    
    \author{Liang-Jun He$^1$}
    
    \author{Fabian Maucher$^2$}
	
	\author{Yong-Chang Zhang$^1$}
	\email{zhangyc@xjtu.edu.cn}
	
	\affiliation{$^1$MOE Key Laboratory for Nonequilibrium Synthesis and Modulation of Condensed Matter, and Shaanxi Key Laboratory of Quantum Information and Quantum Optoelectronic Devices, School of Physics, Xi’an Jiaotong University, Xi’an 710049, People’s Republic of China\\
	$^2$Faculty of Mechanical Engineering; Department of Precision and Microsystems Engineering, Delft University of Technology, 2628 CD, Delft, The Netherlands}
		
\begin{abstract}
    %Quasicrystals form because the underlying interactions feature an incommensurate length-scale competition that inhibits translational periodicity. We show that such a multiscale interaction can be realised in a dipolar Bose-Einstein condensate by the combination of inherent magnetic dipolar interactions and optically-induced interactions by coupling to an excited manifold and a suitably designed feedback, leading to two pronounced roton instabilities in the Bogoliubov excitation spectrum. This interaction leads to a rich phase diagram including quasicrystal ground states that feature twelve-fold rotational symmetry for experimentally realistic parameters. We provide a protocol to access quasicrystals dynamically. Furthermore, we present a general method to calculate the excitation spectrum of quasicrystals. 

Quasicrystals emerge from competing interactions with incommensurate characteristic length scales that inhibit translational periodicity. Here, we show that such multiscale interactions can be realized in a dipolar Bose-Einstein condensate through the interplay between intrinsic dipole-dipole interactions and photon-mediated interactions generated by coupling to an excited-state manifold together with a suitably engineered optical feedback. This interplay gives rise to two pronounced roton instabilities in the Bogoliubov excitation spectrum, leading to a rich ground-state phase diagram including dodecagonal quasicrystals with twelvefold rotational symmetry for experimentally realistic parameters. We further propose a protocol to access these quasicrystals dynamically and develop a general numerical framework for calculating their collective excitation spectra.

\end{abstract}

\maketitle

\emph{Introduction---}The very definition of a crystal was originally based on the assumption of translational periodicity~\cite{Friedrich1912}. 
Unlike periodic crystals, quasicrystals (QCs) cannot be described by a translationally repeating unit cell, yet they exhibit sharp diffraction peaks and may possess noncrystallographic rotational symmetries~\cite{shechtman1984metallic,JassenBook}. QCs can form when interparticle interactions involve multiple competing length scales~\cite{barkan2011stability,archer2013quasicrystalline,barkan2014controlled}
that frustrate translationally periodic lattices, rendering quasiperiodic order the lowest free energy solution. 
QCs were first discovered in Al-Mn alloys in 1984~\cite{shechtman1984metallic}, where the effective pair interaction features short-range and oscillatory longer-range interactions due to Friedel oscillations~\cite{Steurer:QCs:2012}. While both energetic and entropic mechanisms are known~\cite{steurer2018quasicrystals,mivehvar2019emergent} to stabilize QCs, the relative importance of these mechanisms remains system-dependent and their growth mechanism is still an active area of research~\cite{keys2007how,achim2014growth,nagao2015experimental,han2020dynamic,han2021formation,xue2022atomic,de2024mesoscale,xue2025investigation,gao2025direct}. 

Bose-Einstein condensates (BEC) emerged as an ideal platform for exploring ultracold many-body physics~\cite{RevModPhys.80.885} due to the high degree of control and tunability of interatomic interactions~\cite{courteille1998observation,inouye1998observation,theis2004tuning}, including long-range interactions leading to a broad range of crystals with a rich spectrum of symmetries. Such tailored interactions were successfully employed to realize supersolids -- i.e., states with discrete translational symmetry whilst maintaining a large superfluid fraction~\cite{leggett1970can,chester1970pra,leggett1998on} -- by employing additional optical fields~\cite{li2017stripe,leonard2017supersolid} as well as in dipolar Bose-Einstein condensates (dBECs)~\cite{bottcher2019transient,tanzi2019observation,chomaz2019long,norcia2021two}. DBECs~\cite{Bttcher2020,Chomaz2022,pohl:RMP:2026} stand out in that they feature both anisotropic long-range and tunable short-range interactions and stabilize due to quantum fluctuations, %~\cite{Pfau:nature2:2016,Ferlaino:PRX:2016}
which can lead to a rich phase diagram~\cite{zhang2019supersolidity,Hertkorn:PRR:2021,zhang2021phases}. While quasicrystalline order can be directly imprinted onto BECs through carefully designed external quasiperiodic potentials~\cite{viebahn2019matter,niu2020bose,zampronio2024exploring}, the experimental realization of self-organized QCs remains an elusive goal. Theoretically, several proposals have been put forward, including schemes based on spin-orbit coupling~\cite{gopalakrishnan2013quantum,hou2018superfluid} and long-range interactions with multiple characteristic length scales~\cite{dotera2014mosaic,heinonen2019quantum,pupillo2020quantum,campos2021nonconventional,mendoza2022exploring,grossklags2024self,grossklags2025engineering,mendoza2025low}.

\begin{figure}[!t]
	\centering
	\includegraphics[width=0.9\columnwidth]{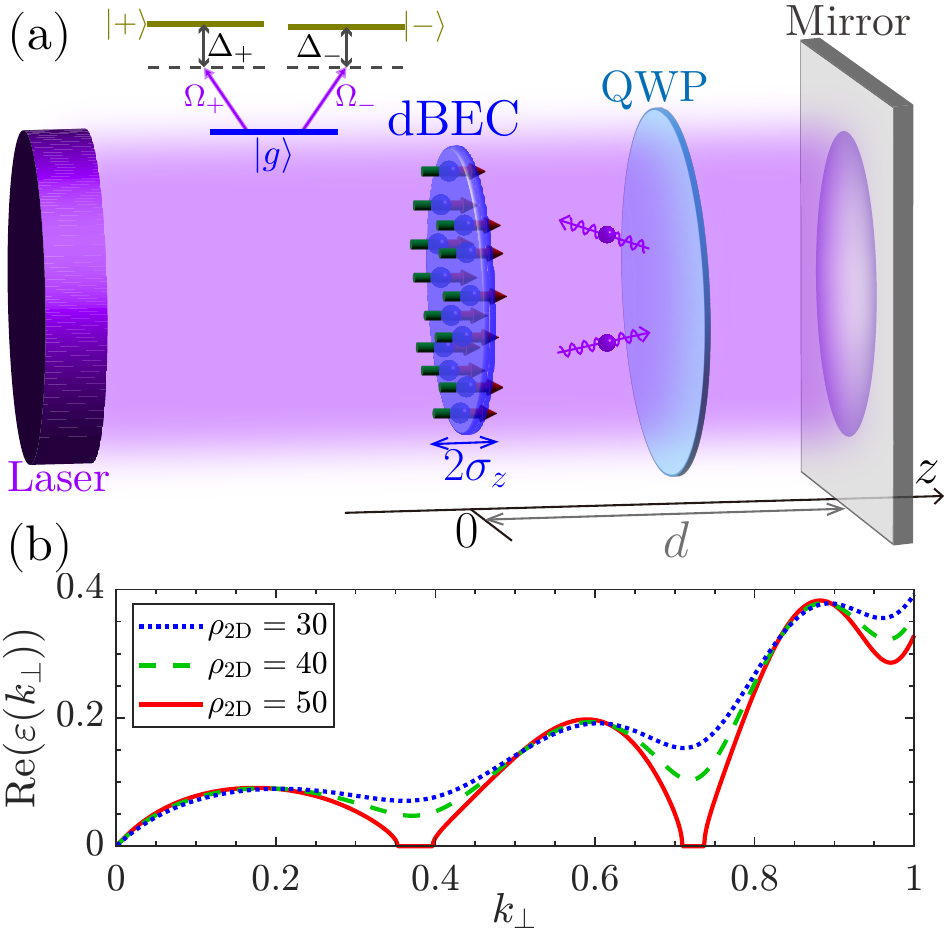}
	\caption{(a) Illustration of the setup. A quasi-2D dBEC polarized along the $z$-direction is placed in front of a retroreflecting mirror at a distance $d$. The thickness of the dBEC is $2\sigma_z$. The forward $\Omega_+$ and backward $\Omega_-$ propagating optical fields off-resonantly couple the ground state to the excited states of the atoms via a V-type configuration with detunings $\Delta_{+}$ and $\Delta_{-}$, respectively. The wavelength of $\lambda=421$nm corresponds to the transition of ${}^{164}$Dy from $|j=8,m_j=-8\rangle$ to $|j'=9,m_j'=-7\rangle$ and $|j'=9,m_j'=-9\rangle$, respectively (cf.~\cite{supply}). (b) A Bogoliubov excitation spectrum for different 2D densities $\rho_{2 \mathrm{D}}$ at $a_s/a_{\mathrm{dd}}=0.75$. The red line corresponds to \textcolor[RGB]{0,120,0}{\large$\bullet$} in Fig.~\ref{phase}. }
	\label{system}
\end{figure}
\nocite{hugenholtz1959ground,pastukhov2017beyond,bisset2021quantum,DivSeries,lu2011spectroscopy,martin1978atomic,Kramida1999-gv} 

In this paper, we provide a setup for realising self-organized dodecagonal quantum QCs in dBECs. For that, we combine the inherent dipole-dipole interactions (DDI) present in dBECs with effective light-induced long-range oscillatory interactions provided by a retroreflecting mirror~\cite{zhang2018long,zhang2021self,zhang2025nonreciprocal}, thereby realising a zero-entropy quantum analogue of classical solid-state alloy physics. We numerically compute the phase diagram and develop a general numerical framework for calculating the excitation spectra of QCs, which, thus far, has been limited to analytical methods involving small wave vectors~\cite{mendoza2025low}. Finally, we design a protocol to dynamically access QCs.

\emph{Theoretical model---}We consider a quasi-two-dimensional (quasi-2D) dBEC located in front of a retroreflecting mirror at a distance $d$ as shown in Fig.~\ref{system}(a). The dBEC is polarized along the $z$ direction and is confined by a strong harmonic trap $U(z)=m\omega_z^2 z^2/2$, where $m$ denotes the atomic mass and $\omega_z$ the trapping frequency. The inherent interactions between the atoms involve short-range repulsion due to $s$-wave scattering with scattering length $a_s$ as well as long-range DDI with a dipolar length $a_{\mathrm{dd}}$. The intrinsic interaction in three-dimensional (3D) space is given by 
\begin{equation}
	V_{\rm atom}(\mathbf{r}-\mathbf{r}')=
	g\delta(\mathbf{r}-\mathbf{r}')+V_{\mathrm{DDI}}(\mathbf{r}-\mathbf{r}'),
\end{equation}
where $g=\frac{4\pi\hbar^2 a_s}{m}$, $V_{\mathrm{DDI}}(\mathbf{r})=\frac{3\hbar^2 a_{\mathrm{dd}}}{m} \left(1-3z^2/r^2\right)$, and $\mathbf{r}=(x,y,z)$. 
%In addition, the dBEC is illuminated by a laser. After traversing the condensate, the light field propagates forward to and is then reflected from the mirror. 
Apart from these inherent interatomic interactions, an additional optical beam couples the atomic ground state to an excited state manifold, leading to an additional long-range interaction. 
The beam passes through the BEC twice due to retroreflection from a mirror. A quarter-wave plate (QWP) reverses the helicity, thus the incident and reflected circularly polarised beams have opposite handedness.
%which gives rise to an additional long-range interaction as follows: 
%Assuming that the incident forward-propagating beam (along the positive $z$-direction) light is circularly polarised (e.g. left-handed), then the reflected backward-propagating beam would feature a reversed the handedness due to the presence of a quarter wave plate (QWP) between the dBEC and the mirror (e.g. right-handed). 
Therefore, the forward and backward propagating beams drive different transitions and couple the atomic ground state to two different excited states $|\pm\rangle$ with detunings $\Delta_+$ and $\Delta_-$, respectively, through a V-type coupling scheme. In the case of large detuning -- i.e., the Rabi-frequency $\Omega_{\pm}={\mu_\pm\mathcal{E}_0}/{\hbar}$ satisfies $|\Delta_\pm|\gg \Omega_\pm$ with $\mathcal{E}_0$ being the amplitude of the incident beam and $\mu_\pm$ the dipole matrix elements -- the optical fields can be adiabatically eliminated, yielding the following effective
optically induced atomic interaction~\cite{zhang2018long,zhang2021self},
\begin{equation}
	V_{\rm opt}\left(\mathbf{r}_{\perp}-\mathbf{r}^{\prime}_{\perp}\right)=\hbar \tilde{\gamma} \cos \left(\frac{k_l}{4 d}\left|\mathbf{r}_{\perp}-\mathbf{r}^{\prime}_{\perp}\right|^2\right),
	\label{V_mirror}
\end{equation}
where $k_l=2\pi/\lambda$ is the wave number of the optical field, $\tilde{\gamma}=-\frac{k^2_l\mu^2_+\mu^2_-\mathcal{E}_0^2}{16\pi \hbar^3 d \epsilon_0 \Delta_{+}\Delta_{-}}$ with $\epsilon_0$ being the vacuum permittivity, 
%$\tilde{\gamma}=-\frac{3\Omega^2\Gamma}{16k_l d \Delta_{+}\Delta_{-}}$, $\Omega=\frac{\mu\mathcal{E}_0}{\hbar}$. Here, $\mathcal{E}_0$ and $\Gamma$ denote the amplitude of the incident beam and the spontaneous decay rate of the excited states, respectively, and we have assumed an equal dipole matrix element $\mu$ for the two transitions. 
and $\mathbf{r}_{\perp}=(x,y)$ denotes the position in the transverse plane. The effective atomic interactions driven by the optical feedback system depend only on the position perpendicular to the mean Poynting flux of the beam. 

%In dBECs, quantum fluctuations may play a significant role, which is usually described by Lee-Huang-Yang (LHY) correction~\cite{lee1957many,lee1957eigenvalues,lima2011quantum,lima2012beyond}. To estimate its contribution in the system considered here, we first consider in three-dimensional (3D) space. The interaction potential with $s$-wave scattering and DDI included is
%\begin{equation}
%	V_{3\mathrm{D}}(\mathbf{r}-\mathbf{r}')=
%	g\delta(\mathbf{r}-\mathbf{r}')+V_{\mathrm{dd},3\mathrm{D}}(\mathbf{r}-\mathbf{r}'),
%\end{equation}
%where $g=\frac{4\pi\hbar^2 a_s}{m}$, $V_{\mathrm{dd},3\mathrm{D}}(\mathbf{r})=\frac{3\hbar^2 %a_{\mathrm{dd}}}{m}
%\left(1-3z^2/r^2\right)$ and $\mathbf{r}=(x,y,z)$ represents the position in 3D space. 

In the presence of these interactions, the energy functional of the dBEC is given by
\begin{equation}
	\begin{split}
	E=\int\mathrm{d}^3\mathbf{r}&\bigg[\Psi^*\left(-\frac{\hbar^2}{2m}\nabla^2\right)\Psi
		+\frac{1}{2}m\omega_z^2 z^2|\Psi|^2\\
		&+\frac{1}{2}g|\Psi|^4
		+\frac{2}{5}\gamma_{3\mathrm{D}}|\Psi|^5
		\bigg]+E_{\mathrm{long}}
	\end{split}
    \label{eq:3}
\end{equation}
where $\Psi(\mathbf{r})$ is the 3D wave function that satisfies $\int|\Psi|^2\mathrm{d}^3 \mathbf{r}=N$ with $N$ being the total number of atoms, $\gamma_{3\mathrm{D}}={\frac{128}{3 m} \sqrt{\pi} \hbar^2 a_s^{{5}/{2}}}(1+\frac{3}{2}(\frac{a_{\mathrm{dd}}}{a_s})^2)$ denotes the coefficient of the Lee-Huang-Yang (LHY) correction~\cite{supply,lee1957many,lee1957eigenvalues,lima2011quantum,lima2012beyond,wachtler2016quantum,baillie2016self,bisset2016ground,wachtler2016ground}, and $E_\mathrm{long}=2^{-1}\iint\mathrm{d}^3\mathbf{r}\mathrm{d}^3\mathbf{r}'
		|\Psi(\mathbf{r})|^2 [V_{\mathrm{DDI}}(\mathbf{r}-\mathbf{r}') +V_{\mathrm{opt}}\left(\mathbf{r}_{\perp}-\mathbf{r}^{\prime}_{\perp}\right)]
		|\Psi(\mathbf{r}')|^2$ represents the contribution of the atomic long-range interactions. 

%as below,
%\begin{equation}
%	\begin{split}
%		E_\mathrm{long}=\frac{1}{2}&\iint\mathrm{d}^3\mathbf{r}\mathrm{d}^3\mathbf{r}'
%		|\Psi(\mathbf{r})|^2 \big[V_{\mathrm{DDI}}(\mathbf{r}-\mathbf{r}') \\ &+V_{\mathrm{mirror}}\left(\mathbf{r}_{\perp}-\mathbf{r}^{\prime}_{\perp}\right)\big]
%		|\Psi(\mathbf{r}')|^2. 
%	\end{split}
%    \label{eq:4}
%\end{equation}
%The coefficient of LHY correction is~\cite{supply}
%\begin{equation}
%	\gamma_{3\mathrm{D}}=\frac{128 \sqrt{\pi} \hbar^2 a_s^{\frac{5}{2}}}{3 m}\left[1+\frac{3}{2}\left(\frac{a_{\mathrm{dd}}}{a_s}\right)^2\right], 
%	\label{gamma3D}
%\end{equation}
%which takes the same form as in usual dipolar gas~\cite{lima2011quantum,lima2012beyond,bisset2016ground,baillie2016self,wachtler2016quantum,wachtler2016ground}. 

We assume that the total density distribution of the condensate can be decomposed as $|\Psi(\mathbf{r})|^2 =|\psi(\mathbf{r}_\perp)|^2\rho_{\rm TF}(z)$.
Since the quasi-2D dBEC is strongly confined along the $z$-axis, we assume a Thomas-Fermi profile $\rho_{\rm TF}(z)=\frac{3}{4 \sigma_z}(1-\frac{z^2}{\sigma_z^2})$~\cite{zhang2019supersolidity,zhang2021phases} within the range of $z\in (-\sigma_z,\sigma_z)$ that is normalized to unity. 
Substituting this ansatz into Eq.~(\ref{eq:3}) and integrating out the axial direction, it is straightforward to find the following effective 2D Gross-Pitaevskii equation (GPE) describing the dynamics of the quasi-2D dBEC,
%Upon rescaling the spatial length and energy with the units of $\ell=12\pi a_{\mathrm{dd}}$ and $\hbar^2/(m \ell^2)$, respectively, the typical width is given by $\sigma_z=(\frac{\rho_{2 \mathrm{D}}\left(a_s / a_{\mathrm{dd}}+2\right)}{2 \omega_z^2})^{1 / 3}$~\cite{zhang2019supersolidity,zhang2021phases,zhang2023variational,he2025accessing}, where $\rho_{2 \mathrm{D}}=N/(\int \mathrm{d}x\mathrm{d}y)$ is the average 2D density.
\begin{align}
	%\begin{split}
		i\frac{\partial\psi}{\partial t}=&\bigg\{-\frac{\nabla^2_\perp}{2}
		+g_{2\mathrm{D}}|\psi|^2
		+\gamma_{2\mathrm{D}}|\psi|^3 + \int\big[U^{\rm 2D}_{\mathrm{DDI}}(\mathbf{r}_{\perp}-\mathbf{r}_{\perp}') \nonumber\\
        &+U_{\mathrm{opt}}(\mathbf{r}_{\perp}-\mathbf{r}_{\perp}')\big]|\psi(\mathbf{r}_\perp')|^2 \mathrm{d}^2\mathbf{r}_{\perp}'
		\bigg\}\psi.
	%\end{split}
    \label{GPE_2D}
\end{align}
Here, $\psi(\mathbf{r}_{\perp})$ is the 2D wave function normalized to the total particle number, i.e., $\int|\psi(\mathbf{r}_{\perp})|^2\mathrm{d}^2\mathbf{r}_{\perp}=N$. The GPE has been non-dimensionalized by rescaling the spatial length and energy with the units of $\ell=12\pi a_{\mathrm{dd}}$ and $\hbar^2/(m \ell^2)$, respectively, leading to the dimensionless parameters  $g_{2\mathrm{D}}=\frac{a_s}{5\sigma_z a_{\mathrm{dd}}}$, $\gamma_{2\mathrm{D}}=\frac{15\sqrt{3}}{128\pi\sigma_z^{{3}/{2}}} \left(\frac{a_s}{3a_{\mathrm{dd}}}\right)^{{5}/{2}}	\left[1+\frac{3}{2}\left(\frac{a_{\mathrm{dd}}}{a_s}\right)^2\right]$, and the dimensionless 2D interactions $U^{\rm 2D}_{\mathrm{DDI}}(\mathbf{r}_{\perp})$ and $U_{\mathrm{opt}}(\mathbf{r}_{\perp})$. Hereafter, our discussion will be conducted in these dimensionless units at a fixed trapping frequency of $\omega_z=0.15$. The typical width is approximately $\sigma_z=(\frac{\rho_{2 \mathrm{D}}\left(a_s / a_{\mathrm{dd}}+2\right)}{2 \omega_z^2})^{1 / 3}$~\cite{zhang2019supersolidity,zhang2021phases} with $\rho_{2 \mathrm{D}}$ being the average 2D density.
%$\rho_{2 \mathrm{D}}=N/(\int \mathrm{d}x\mathrm{d}y)$ being the average 2D density.

%The convolution kernel $U_{\mathrm{dd}}(\mathbf{r})$ and $U_{\mathrm{m}}(\mathbf{r})$ represent effective 2D DDI and interaction induced by light field respectively, and their Fourier transformations are
%\begin{equation}
%		\tilde{U}_{\mathrm{dd}}(\mathbf{k}_\perp)=\frac{3}{4 \sigma_z}\left(f(k_{\perp} \sigma_z)-\frac{4}{15}\right)
%		\label{U_dd}
%\end{equation}
%and
%\begin{equation}
%	\tilde{U}_{\mathrm{m}}(\mathbf{k}_\perp)=\alpha\sin(\beta k^2),
%	\label{U_m}
%\end{equation}
%with $f(x)=[3-3 x^2+2 x^3-3\left(1+x\right)^2 e^{-2 x}]/x^5$, where $\alpha=\frac{2\tilde{\gamma}dm}{\hbar k_l}$ and $\beta=\frac{d}{k_l \ell^2}$. Eq.~\eqref{U_dd} and Eq.~\eqref{U_m} can be used to simplify the convolution calculation in Eq.~\eqref{GPE_2D}. 

%Hereafter, we will consider in 2D space and fix $\omega_z=0.15$. 

To investigate the symmetry-breaking behavior of this dBEC driven by optical fields, we first examine the Bogoliubov excitation spectrum~\cite{ozeri2005colloquium} of a homogeneous superfluid with density $\rho_{\rm 2D}$, which is given by
%For the homogeneous superfluid state with density $n=|\psi(\mathbf{r}_{\perp})|^2=\rho_{2 \mathrm{D}}$, its elementary excitations can be described by Bogoliubov spectrum~\cite{ozeri2005colloquium}, which yields
\begin{equation}
	\varepsilon(k_{\perp})=\sqrt{\frac{k_{\perp}^2}{2}\left(\frac{k_{\perp}^2}{2}+2\rho_{2 \mathrm{D}}\tilde{U}_{\mathrm{int}}(k_\perp)+3\gamma_{2\mathrm{D}}\rho_{2 \mathrm{D}}^{\frac{3}{2}}\right)}.
\label{eq:bogoliubov}
\end{equation}
Here, $\tilde{U}_{\mathrm{int}}(\mathbf{k}_\perp)=g_{\rm 2D}+\tilde{U}^{\rm 2D}_{\mathrm{DDI}}(\mathbf{k}_\perp)+\tilde{U}_{\mathrm{opt}}(\mathbf{k}_\perp)$, $\tilde{U}^{\rm 2D}_{\mathrm{DDI}}(\mathbf{k}_\perp)=\frac{3f(k_{\perp} \sigma_z)}{4 \sigma_z}-\frac{1}{5\sigma_z}$, $\tilde{U}_{\mathrm{opt}}(\mathbf{k}_\perp)=\alpha\sin(\beta k^2_\perp)$, with $\alpha=\frac{2\tilde{\gamma}dm}{\hbar k_l}$, $\beta=\frac{d}{k_l \ell^2}$, and $f(x)=[3-3 x^2+2 x^3-3\left(1+x\right)^2 e^{-2 x}]/x^5$. As shown in Fig.~\ref{system}(b), by a suitable choice of  parameters, we can realise multiple roton instabilities and the ratio between the roton momenta is highly tunable. If the ratio of the two unstable-roton momenta in the red line is approximately $\sqrt{2+\sqrt{3}}$, it promotes the stabilization of dodecagonal QCs~\cite{grossklags2025engineering,pupillo2020quantum}.

\emph{Ground states---}We explore the ground states in the thermodynamic limit, where both the total number of atoms $N$ and the area $\mathcal{A}$ diverge, while the average 2D density $\rho_{2 \mathrm{D}}=N/\mathcal{A}$ maintains a finite value. We find the ground states numerically by propagating Eq.~(\ref{GPE_2D}) in imaginary time with periodic boundary conditions (PBC). There is a subtlety related to the fact that QCs are not periodic, such that there is an inevitable impact imposed by PBC. Thus, it is crucial to reassert that the found QCs are actual ground states of the system rather than a numerical artifact (cf.~\cite{supply}). 
\begin{figure}[!b]
	\centering
	\includegraphics[width=\columnwidth]{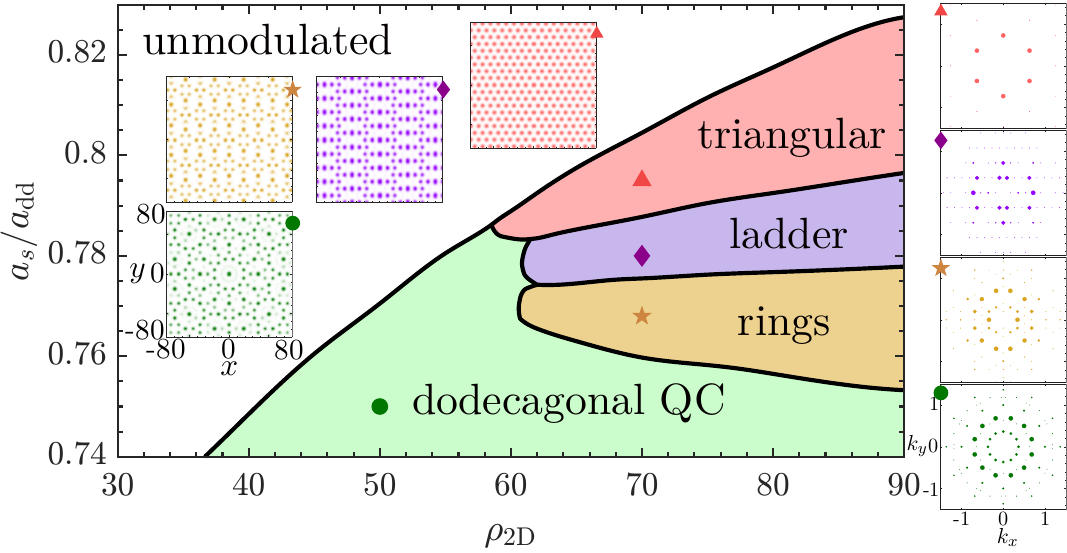}
	\caption{Ground-state phase diagram as a function of $\rho_{2 \mathrm{D}}$ and $a_s/a_{\mathrm{dd}}$ at $\alpha=-0.00148$ and $\beta=14.69$. The subgraphs present the density distributions of the four modulated states in the same colors, which are obtained using the parameters $\rho_{2 \mathrm{D}}$ and $a_s/a_{\mathrm{dd}}$ marked by \textcolor[RGB]{0,120,0}{\large$\bullet$}, \textcolor[RGB]{186,142,35}{$\filledstar$}, \textcolor[RGB]{139,0,139}{$\blacklozenge$}, and \textcolor[RGB]{240,70,70}{$\blacktriangle$}, respectively. The right column displays their corresponding momentum profiles.}
	\label{phase}
\end{figure}

Fig.~\ref{phase} exhibits the ground-state phase diagram at fixed values of $\alpha=-0.00148$ and $\beta=14.69$, the choice of which is discussed later and in~\cite{supply}. There are four different modulated phases. In particular, a dodecagonal QC  represents the ground state of a large parameter domain. The density distribution of this dodecagonal QC state appears to feature no translational periodicity in real space. Moreover, the momentum profile presents a clear twelve-fold rotational symmetry, which is forbidden in periodic crystals~\cite{kittel2004introduction}. At small $\rho_{\rm 2D}$, the system directly transits from the superfluid to the dodecagonal QC state through a first-order phase transition. At higher $\rho_{2 \mathrm{D}}$, three regular crystal phases emerge between the superfluid and the dodecagonal QC state, among which the triangular state features a six-fold rotational symmetry, while the other two states (i.e., the ladder and rings states) both feature a two-fold rotational symmetry. The phase transitions between these states are all of first order as well. 

A qualitative understanding of the phase diagram can be acquired through the Bogoliubov excitation spectrum: The interaction ratio $a_{\rm s}/a_{\rm dd}$ and the density $\rho_{\rm 2D}$ tune the system from a regime with a single pronounced roton minimum, which favors a triangular lattice, to a regime with two comparably deep roton minima, which gives rise to either supercrystals (ladder and rings) or quasicrystals, depending on the ratio of the corresponding roton momenta. As mentioned in the introduction, unlike, e.g., classical alloys, dodecagonal QC in this dipolar quantum gas can exhibit an appreciable superfluid fraction (cf.~\cite{supply}) and excitations without classical counterparts which we explore in what follows.

\emph{Excitation spectrum of the dodecagonal QC---}Unlike periodic crystals, whose excitation spectra can be efficiently computed using Bloch's theorem within a unit cell~\cite{kittel2004introduction,poli2024excitations,blakie2025dirac,cook2026excitations}, QCs lack translational periodicity, rendering Bloch-based approaches inapplicable. We present a general numerical framework for calculating the excitation spectrum of dodecagonal QC~\cite{supply}, by employing the projection method~\cite{jiang2014numerical} which maps a 2D dodecagonal quasicrystal onto a periodic crystal in 4D space. 

\begin{figure}[!t]
	\centering
	\includegraphics[width=1\columnwidth]{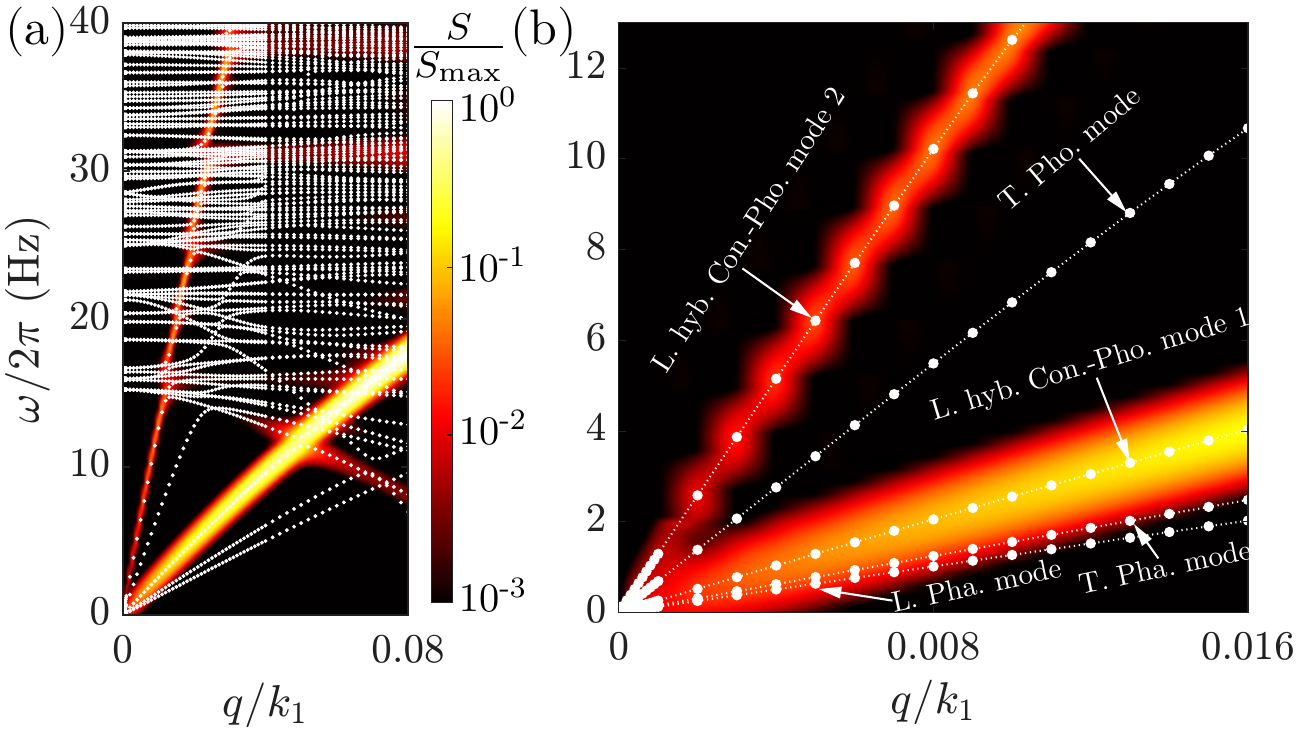}
	\caption{(a) The excitation spectrum and dynamic structure factor of the dodecagonal QC marked by \textcolor[RGB]{0,120,0}{\large$\bullet$} in Fig.~\ref{phase}. (b) Enlarged view of the five gapless modes. The dynamic structure factor is represented by the heat map, with a frequency broadening of $0.005\omega_z$. Here, the momentum $\mathbf{q}$ is taken along $x$-axis.}
	\label{spectrum_8}
\end{figure}

Fig.~\ref{spectrum_8} shows the low-momentum excitation spectrum of the dodecagonal QC state marked by \textcolor[RGB]{0,120,0}{\large$\bullet$} in Fig.~\ref{phase}. We find five gapless modes in agreement with the general theory that 
an $n$-dimensional crystal possesses $n+1$ Goldstone modes  
associated with the spontaneous translational symmetry breaking ($n$) and global U(1) gauge symmetry (1) as well as with~\cite{mendoza2025low}: One longitudinal phason (L.~Pha.) mode, one transverse phason (T.~Pha.) mode, two longitudinal hybrid condensate-phonon modes (L.~hyb.~Con.-Pho.~modes 1 and 2), and one transverse phonon (T.~Pho.) mode. It is worth to note that the frequency ordering of these modes reflected by their sound velocities differs from that found in Ref.~\cite{mendoza2025low}, illustrating the sensitivity of the excitation spectrum to the specific form of the interparticle interactions~\cite{poli2024excitations}.

Furthermore, our approach also yields the dynamical structure factor $S$~\cite{poli2024excitations,BECandSF,Exc2019Ferlaino,blakie2025dirac}. Notably, only the two L.~hyb.~Con.-Phon.~modes possess finite spectral weight, while the other three gapless modes have vanishing dynamical structure factors. This reflects their distinct responses to perturbations with momentum $\mathbf{q}$ in the physical space of the 2D QCs. The two phason modes involve fluctuations perpendicular to the physical plane, while the T.~Pho.~mode corresponds to in-plane fluctuations transverse to the momentum transfer $\mathbf{q}$. As a result, none of these modes couples to density fluctuations probed at momentum $\mathbf{q}$, leading to vanishing dynamical structure factors. In contrast, the two L.~hyb.~Con.-Pho.~modes are characterized by longitudinal density oscillations and, therefore, exhibit finite dynamical responses.

%Additionally, our approach also provides information on the dynamical structure factor $S$. Notably, only the two L.~hyb.~Con.-Phon.~modes are accompanied with finite $S$, while the other three gapless modes feature vanishing dynamical structure factors, indicating distinct behaviors of these gapless modes in response to perturbations with momentum $\mathbf{q}$ in the physical space of the 2D QCs. Specifically, the T.~Pho. mode and the two phason modes correspond to vibrations in the direction perpendicular to the 2D physical space, and such a constraint suppresses the dynamical response of these modes, while the two L.~hyb.~Con.-Phon.~modes can present global density fluctuations caused by perturbations. 

Our method can also, to a certain degree, resolve the dense spectrum of high-energy excitations~\cite{DenseSpec1,DenseSpec2} beyond the low-energy gapless modes. As shown in Fig.~\ref{spectrum_8}(a), the two L.~hyb.~Con.-Pho.~modes extend smoothly into high-frequency branches. With sufficient computational resources, excitation spectra at large momenta far beyond the linear regime can be computed as well.

\begin{figure}
    \includegraphics[width=0.9\columnwidth]{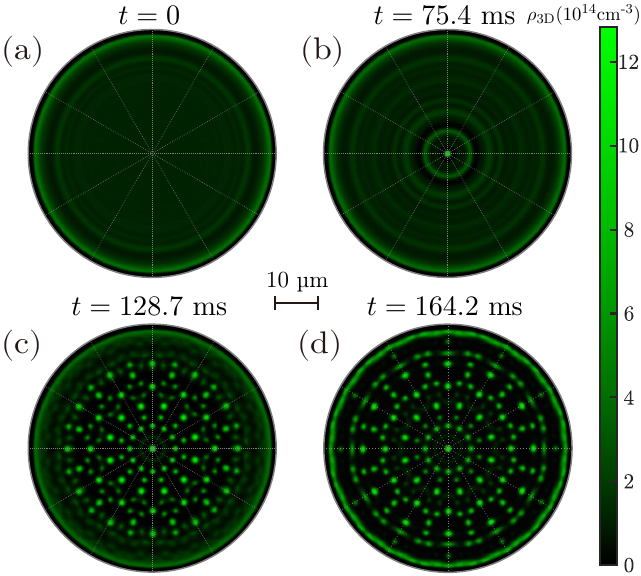}
	\caption{Dynamical evolution of the dBEC driven by the local and gradual quench [cf. Eqs.~(\ref{eq:box}) and~(\ref{eq:gradual_scattering})] with $v_0=0.4$.}
	\label{dynamics_3bl}
\end{figure}

\emph{Spontaneous formation of the dodecagonal QC---}We proceed with exploring the real-time excitation of QCs and present a scheme to dynamically access the  dodecagonal QC state. For that, we consider a dBEC confined in the following box trap with a radius of $R$ in the transverse direction, 
\begin{equation}
	U_{\mathrm{box}}(r_\perp)=\begin{cases} 0, & \mathrm{ if } ~ r_{\perp} \leqslant R \\ \frac{1}{2}\omega_{\perp}^2\left(r_{\perp}-R\right)^2, & \mathrm{if} ~ r_{\perp}>R\end{cases}, 
    \label{eq:box}
\end{equation}
i.e., a constant potential in the center and a harmonic potential with the frequency of $\omega_\perp$ for $r_{\perp}>R$. In the numerical simulation, we have fixed $R=100$, $\omega_\perp=1$, and $\rho_{\rm 2D}=50$. 

The condensate is initialized to the unmodulated ground state at $a_s/a_{\mathrm{dd}}=0.85$ in this box potential and we choose the other parameter as in Fig.~\ref{phase}. Subsequently, we locally tune the $s$-wave scattering length to $a_s/a_{\mathrm{dd}}=0.75$ in the following manner, 
\begin{equation}
	\frac{a_s(r_{\perp},t)}{a_{\rm dd}}=\begin{cases}0.75, & \mathrm{ if } ~ r_{\perp} \leqslant v_0 t \\ 0.85, & \mathrm{if} ~ r_{\perp}>v_0 t\end{cases}.
    \label{eq:gradual_scattering}
\end{equation}
That is, instead of a global quench, the reduction of the interaction strength $a_s/a_{\mathrm{dd}}$ (i.e., $0.85\xrightarrow{}0.75$) is gradually conducted from the center to the edge with a diffusion velocity of $v_0$. This can be implemented employing the local Feshbach resonance technique~\cite{Chien2012,PhysRevLett.115.155301}. The contact interaction coefficient $g_{\rm 2D}=a_s(r_\perp,t)/(5\sigma_z a_{\rm dd})$ becomes position- and time-dependent during the dynamics governed by the GPE~(\ref{GPE_2D}) which includes the additional box potential $U_{\rm box}(r_\perp)$ and a three-body loss $-iL^{\rm 2D}_{3}|\psi|^4/2$~\cite{zhang2021phases,bottcher2019dilute,he2025accessing}. Here, $L^{\rm 2D}_{3}=\frac{27m}{70\hbar\ell^4\sigma_z^2}L_3$ represents effective three-body loss rate in quasi-2D space, with $L_3=1.5\times 10^{-41}~\mathrm{m}^6/\mathrm{s}$~\cite{bottcher2019dilute}.

Fig.~\ref{dynamics_3bl}(a)-(d) shows the dynamics of the condensate driven by the above local and gradual quench of the $s$-wave scattering length. At the early stage, only the central atoms enter the dodecagonal QC regime with a weak interaction $a_s/a_{\mathrm{dd}}=0.75$, while the rest remains in the superfluid regime with a strong interaction $a_s/a_{\mathrm{dd}}=0.85$. Therefore, the symmetry breaking starts to occur from the trap center, where the atoms first reshape their distribution into a multi-ring structure [cf. Fig.~\ref{dynamics_3bl}(b)]. Upon further evolution, the rings spontaneously break the continuous rotational symmetry into the twelve-fold rotational symmetry, a typical signature of the dodecagonal QCs. As time progresses, the rotational symmetry breaking proceeds outwards [cf. Fig.~\ref{dynamics_3bl}(c)]. Eventually, the entire system evolves into a dodecagonal QC as displayed in Fig.~\ref{dynamics_3bl}(d).

It is important to note that the local and gradual quench approach designed here is essential to robustly generate a nearly-perfect QC, while it remains challenging to produce a single large-size QC even in classical materials~\cite{ProduceQC1,ProduceQC2}. A global sudden quench would not permit their realisation (cf.~\cite{supply}). 

For the experimental implementation, we provide the relevant parameters used in Fig.~\ref{dynamics_3bl} in real units as follows: $\omega_z=845\mathrm{Hz}$, $\sigma_z=3.8\upmu\mathrm{m}$, $\Delta_{\pm}= 1.82\mathrm{GHz}$ and the pumping intensity $I_{0}=2.32\mathrm{mW}\cdot\mathrm{cm}^{\text{-}2}$. Here, we used $^{164}{\rm Dy}$ as an example and selected $4\mathrm{f}^{10}(^{5}\mathrm{I}_8)6\mathrm{s}6\mathrm{p}(^{1}\mathrm{P}_1)$ as excited states $|\pm\rangle$. To match the transition frequency between the atomic ground state and the excited states, the wavelength of light is $\lambda=421\mathrm{nm}$ (cf. the color of the coupling laser in Fig.~\ref{system}). The effective dBEC-mirror separation is set at $d=14.97\upmu\mathrm{m}$, which can be tuned over a wide range and made much smaller than the actual physical separation~\cite{Ackemann2014NatPhoton}. Alternatively, one can also use the transition channel $4\mathrm{f}^{12}(^{3}\mathrm{H}_6)6\mathrm{s}6\mathrm{p}(^{1}\mathrm{P}_1)$ of $^{166}{\rm Er}$. Further details can be found in the Supplemental Material~\cite{supply}.

\emph{Conclusions---}We propose a scheme to engineer tunable long-range interactions in quasi-2D dBECs via optical feedback. Their interplay with the magnetic DDI produces tunable multiple roton minima, yielding a rich phase diagram that includes dodecagonal QCs. We also introduce a general method for computing QC excitation spectra and a locally gradual quench protocol for dynamically realizing these states. Our results establish a versatile platform for exploring the properties and excitations of quantum QCs.
There are several features that distinguish the presented dodecagonal QCs from QCs found in alloys and other classical materials, such as their finite superfluid fraction and their non-classical excitation spectra. 
Additionally, the setup offers broad tunability: Beyond what was considered here, adjusting, e.g., detunings or the BEC-mirror separation allows further control of the roton minima and with it may access to diverse other unusual ground states, such as kagome~\cite{yin2022topological},  $\text{Ta}_{\text{21}}\text{Te}_{\text{13}}$\cite{conrad2000hexagonal}, other super-lattice states and QCs with ten-fold or other rotational symmetries. Moreover, the scheme can be extended to other long-range interacting systems, such as Rydberg-dressed BECs, opening a versatile platform for studying novel quantum phases arising from competing interactions and length scales.

\emph{Acknowledgments---}We are grateful to Thomas Pohl for helpful discussions. This work was supported by the Quantum Science and Technology-National Science and Technology Major Project (Grant No. 2024ZD0300600), National Nature Science Foundation of China (Grant No. 12104359), Shaanxi Academy of Fundamental Sciences (Mathematics, Physics) (Grant No. 22JSY036). Y.C.Z. acknowledges the support of Xiaomi Young Talents program, Xi'an Jiaotong University through the ``Young Top Talents Support Plan" and Basic Research Funding as well as the High-performance Computing Platform of Xi'an Jiaotong University for the computing facilities.

\bibliography{mybib}

\end{document}

\title{Supplemental Material for ``Self-organized quasicrystals and their excitations in dipolar Bose-Einstein condensates via optical feedback"}
    
    \author{Liang-Jun He$^1$}
    
    \author{Fabian Maucher$^2$}
	
	\author{Yong-Chang Zhang$^1$}
	\email{zhangyc@xjtu.edu.cn}
	
	\affiliation{$^1$MOE Key Laboratory for Nonequilibrium Synthesis and Modulation of Condensed Matter, and Shaanxi Key Laboratory of Quantum Information and Quantum Optoelectronic Devices, School of Physics, Xi’an Jiaotong University, Xi’an 710049, People’s Republic of China\\
	$^2$Faculty of Mechanical Engineering; Department of Precision and Microsystems Engineering, Delft University of Technology, 2628 CD, Delft, The Netherlands}

\maketitle

\onecolumngrid

\vspace{-10mm}

\onecolumngrid

\renewcommand{\thetable}{S\arabic{table}}
\renewcommand{\theequation}{S\arabic{equation}}
\renewcommand{\thefigure}{S\arabic{figure}}

\section{I. Derivation of the LHY correction}
In this section, we present the detailed derivation of the LHY correction used in the main text. In 3D space, the interaction potential consisting of the short-range $s$-wave scattering and the long-range dipole-dipole interaction (DDI) can be written as
\begin{equation}
	V_{\rm atom}(\mathbf{r}-\mathbf{r}')=
	g\delta(\mathbf{r}-\mathbf{r}')+V_{\rm DDI}(\mathbf{r}-\mathbf{r}'),
\end{equation}
and, by performing a Fourier transformation, can be conveniently expressed in momentum space as
\begin{equation}
	\tilde{V}_{\rm atom}(\mathbf{k})=g+\frac{4\pi\hbar^2 a_{\mathrm{dd}}}{m}\left(3\frac{k_z^2}{k^2}-1\right), 
\end{equation}
where $\mathbf{k}=(k_x,k_y,k_z)$ represents the wave vector. 
The effective interaction induced by the optical field is given by~\cite{zhang2018long,zhang2021self}
\begin{equation}
	V_{\rm opt}\left(\mathbf{r}_{\perp}-\mathbf{r}^{\prime}_{\perp}\right)=\hbar \tilde{\gamma} \cos \left(\frac{k_l}{4 d}\left|\mathbf{r}_{\perp}-\mathbf{r}^{\prime}_{\perp}\right|^2\right), 
\end{equation}
with $k_l$ being the wavenumber of the light, $d$ the distance between the condensate and the mirror, and $\tilde{\gamma}$ the effective interaction strength. In momentum space, it reads
\begin{equation}
	\tilde{V}_{\rm opt}(\mathbf{k})=
	\frac{4\pi\hbar\tilde{\gamma}d}{k_l}\sin\left(\frac{d}{k_l}\left(k_x^2+k_y^2\right)\right)\delta(k_z). 
\end{equation}

According to the Hugenholz-Pines formalism~\cite{hugenholtz1959ground,pastukhov2017beyond,bisset2021quantum}, the LHY energy density $\epsilon_{\mathrm{LHY}}$ satisfies the following  differential equation: 
\begin{equation}
    \epsilon_{\mathrm{LHY}}-\frac{\rho}{2}\frac{\partial\epsilon_{\mathrm{LHY}}}{\partial \rho}=
    -\frac{1}{8}\int\frac{\mathrm{d}^3\mathbf{k}}{(2\pi)^3}\frac{\left(\varepsilon_{\mathbf{k}}-\tau_{\mathbf{k}}\right)^3}{\varepsilon_{\mathbf{k}}\tau_{\mathbf{k}}}, 
    \label{HP}
\end{equation}
where $\varepsilon_{\mathbf{k}}=\sqrt{\tau_{\mathbf{k}}[\tau_{\mathbf{k}}+2\rho(\tilde{V}_{\rm atom}(\mathbf{k})+\tilde{V}_{\rm opt}(\mathbf{k}))]}$ with $\tau_{\mathbf{k}}=\frac{\hbar^2 k^2}{2m}$. 
By rescaling the momentum as $\mathbf{k}\xrightarrow{} \sqrt{8\pi \rho a_s}\tilde{\mathbf{k}}$, the right hand side of Eq.~\eqref{HP} can be expressed as $G\frac{\hbar^2}{m}(\rho a_s)^{\frac{5}{2}}$, where $G=-\sqrt{\frac{2}{\pi}}\int\mathrm{d}^3\tilde{\mathbf{k}}
\frac{\left(\sqrt{\tilde{k}^2\left(\tilde{k}^2+\frac{m}{2\pi\hbar^2 a_s}\left(\tilde{V}_{\mathrm{atom}}(\tilde{k})+\tilde{V}_{\mathrm{opt}}(\tilde{k})\right)\right)}-\tilde{k}^2\right)^3}{\tilde{k}^2\sqrt{\tilde{k}^2\left(\tilde{k}^2+\frac{m}{2\pi\hbar^2 a_s}\left(\tilde{V}_{\mathrm{atom}}(\tilde{k})+\tilde{V}_{\mathrm{opt}}(\tilde{k})\right)\right)}}$ is a dimensionless constant. Based on dimensional analysis, it can be assumed that the LHY correction takes the form of $\epsilon_{\mathrm{LHY}}=\tilde{G} \frac{\hbar^2}{m}(\rho a_s)^{\frac{5}{2}}$ where $\tilde{G}$ is a dimensionless constant. This immediately yields $\rho\frac{\partial\epsilon_{\mathrm{LHY}}}{\partial \rho}=\frac{5}{2}\epsilon_{\mathrm{LHY}}$. Substituting this relation into Eq.~\eqref{HP}, we  obtain the LHY correction as follows, 
\begin{equation}
	\epsilon_{\mathrm{LHY}}=
	\frac{1}{2}\int\frac{\mathrm{d}^3\mathbf{k}}{(2\pi)^3}\frac{\left(\varepsilon_{\mathbf{k}}-\tau_{\mathbf{k}}\right)^3}{\varepsilon_{\mathbf{k}}\tau_{\mathbf{k}}}.
    \label{eq:S6}
\end{equation}

To carry out the integration over $\mathbf{k}$ in the above equation, we divide the integration domain into two regions: a low-momentum regime $|k_z|<\eta$ near $k_z=0$ and the complementary region of $|k_z|>\eta$. Consequently, Eq.~(\ref{eq:S6}) can be decomposed as  
\begin{equation}
	\begin{split}
		\epsilon_{\mathrm{LHY}}=& \frac{1}{2(2\pi)^3}\lim\limits_{\eta\to 0^{+}}
		\int_{0}^{+\infty}\mathrm{d}k_{\perp}\left(\int_{-\eta}^{\eta}\mathrm{d}k_z+\int_{-\infty}^{-\eta}\mathrm{d}k_z+\int_{\eta}^{+\infty}\mathrm{d}k_z\right)
		k_{\perp}\frac{\left(\varepsilon_{\mathbf{k}}-\tau_{\mathbf{k}}\right)^3}{\varepsilon_{\mathbf{k}}\tau_{\mathbf{k}}}\\
		=& \frac{1}{2}\int\frac{\mathrm{d}^3\mathbf{k}}{(2\pi)^3}\frac{\left(\varepsilon'_{\mathbf{k}}-\tau_{\mathbf{k}}\right)^3}{\varepsilon'_{\mathbf{k}}\tau_{\mathbf{k}}}
		+\frac{1}{2(2\pi)^3}\lim\limits_{\eta\to 0^{+}}\int_{0}^{+\infty}\mathrm{d}k_{\perp}\int_{-\eta}^{\eta}\mathrm{d}k_z
		~k_{\perp}\frac{\varepsilon_{\mathbf{k}}^2}{\tau_{\mathbf{k}}}\left(1-\frac{\tau_{\mathbf{k}}}{\varepsilon_{\mathbf{k}}}\right)^3, 
		\label{ELHY_int}
	\end{split}
\end{equation}
with $\varepsilon'_{\mathbf{k}}=\sqrt{\tau_{\mathbf{k}}[\tau_{\mathbf{k}}+2\rho\tilde{V}_{\rm atom}(\mathbf{k})]}$. Since the integral in the first term of Eq.~\eqref{ELHY_int} does not include the point of $k_z=0$, where $\tilde{V}_{\rm opt}(\mathbf{k})$ vanishes, the excitation frequency $\varepsilon'_{\mathbf{k}}$ depends only on the DDI between the atoms. Therefore, this term reduces to the usual LHY correction in bare dBECs, i.e.,
\begin{equation}
	\begin{split}
		\epsilon_{\mathrm{LHY},1}&= \frac{16 \sqrt{2\pi} \hbar^2 (\rho a_s)^{\frac{5}{2}}}{m}\int_{0}^{+\infty}\mathrm{d}q\int_{0}^{1}\mathrm{d}u
		~q^2 \frac{\left(\sqrt{q^2+2f(u)}-q\right)^3}{\sqrt{q^2+2f(u)}}\\
		&=\frac{256 \sqrt{\pi} \hbar^2 (\rho a_s)^{\frac{5}{2}}}{15 m}\mathcal{Q}_5\left(\frac{a_{\mathrm{dd}}}{a_s}\right)
	\end{split}
\end{equation}
where $f(u)=1+\frac{a_{\mathrm{dd}}}{a_s}(3u^2-1)$, and $\mathcal{Q}_5\left(\frac{a_{\mathrm{dd}}}{a_s}\right)=\int_{0}^{1}f^{\frac{5}{2}}(u)\mathrm{d}u\approx 1+\frac{3}{2}\left(\frac{a_{\mathrm{dd}}}{a_s}\right)^2$~\cite{baillie2016self,bottcher2019dilute,bisset2016ground,wachtler2016quantum,wachtler2016ground}. For the second term in Eq.~\eqref{ELHY_int}, $\tilde{V}_{\rm opt}(\mathbf{k})$ diverges as $k_z\xrightarrow{}0$. Hence, it is justified to approximate $\varepsilon_{\mathbf{k}}\approx \sqrt{2\rho\tau_{\mathbf{k}}\tilde{V}_{\rm opt}(\mathbf{k})}$, which yields
\begin{equation}
	\begin{split}
		\epsilon_{\mathrm{LHY},2}&=\frac{\rho}{(2\pi)^3}\lim\limits_{\eta\to 0^{+}}\int_{0}^{+\infty}\mathrm{d}k_{\perp}\int_{-\eta}^{\eta}\mathrm{d}k_z
		~k_{\perp}\tilde{V}_{\rm opt}(\mathbf{k})\\
		&=\frac{\hbar\tilde{\gamma}d \rho}{2\pi^2 k_l}\int_{0}^{+\infty}\mathrm{d}k_{\perp}~k_{\perp}\sin\left(\frac{d}{k_l}k_{\perp}^2\right)\\
		&=\frac{\hbar\tilde{\gamma}\rho}{4\pi^2}\lim\limits_{\xi\to 0^{+}}
		\int_{0}^{+\infty}\mathrm{d}q~\mathrm{e}^{-\xi q}\sin(q)\\
		&=\frac{\hbar\tilde{\gamma}\rho}{4\pi^2}. 
	\end{split}
\end{equation}
To evaluate the above integral, we adopt a regularization procedure~\cite{DivSeries}. The reulsting term $\epsilon_{\mathrm{LHY},2}$ is proportional to the density $\rho$ and can therefore be discarded, since its contribution to the total chemical potential $\frac{\partial}{\partial \rho}\epsilon_{\mathrm{LHY},2}$ is a density-independent constant and therefore does not influence the GPE. That is, the light-induced interaction leaves the the LHY correction of the dBECs unchanged. Consequently, the LHY energy density is $\epsilon_{\mathrm{LHY}}=\frac{2}{5}\gamma_{3\mathrm{D}}\rho^{\frac{5}{2}}$ [see Eq.~(3) in the main text].

%It should be noted that in realistic 3D systems, the light field causes interactions between atoms along the $z$ direction as well. However, our approach remains reasonable. Since we will ultimately consider in quasi-2D circumstance, even if there is an interaction caused by the light field along the $z$ direction, it can be neglected. 
%Furthermore, even if we take such interaction into consideration, it will ultimately only affect the coefficient of the LHY correction. Such a change will at most cause differences in the results quantitatively but not qualitatively~\cite{he2024infrared}.

\section{II. Numerical approach for identifying quasicrystalline states}

\begin{figure}[!t]
	\centering
	\includegraphics[width=1\columnwidth]{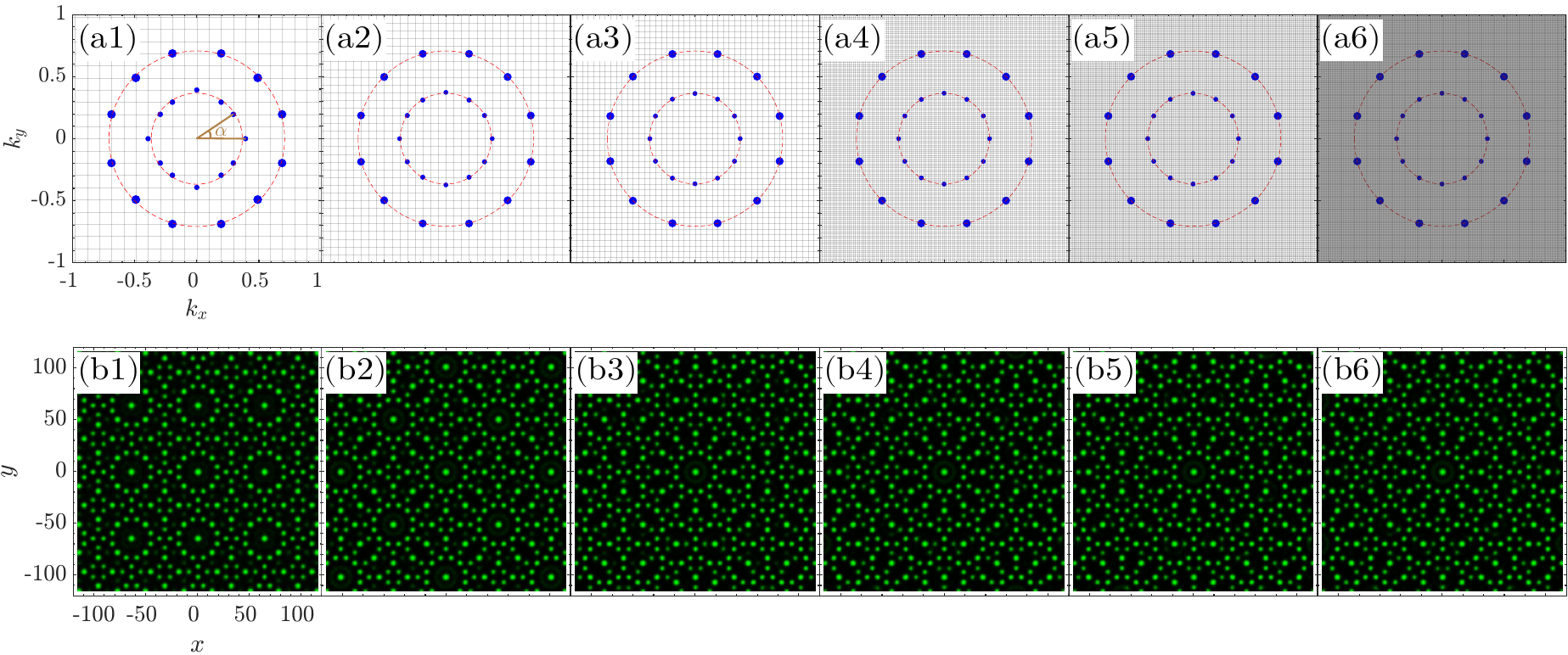}
	\caption{The convergence process of a dodecagonal QC calculated using square grids with increasing resolution in the momentum space. (a1$\sim$a6) present the main components in momentum space, where the zero-momentum peak has been omitted. (b1$\sim$b6) display the respective density distributions in real space. From left to right, the corresponding $\alpha$ has been fixed at $\tan\alpha=2/3,\ 3/5,\ 4/7,\ 11/19,\ 15/26$, and 41/71, respectively. The rest relevant parameters are $\rho_{2 \mathrm{D}}=50$, $a_s/a_{\mathrm{dd}}=0.75$, $\alpha=-0.00148$ and $\beta=14.69$.}
	\label{bijin}
\end{figure}

In this section, we provide additional details on the numerical calculation of quasicrystalline states. In our scheme, the dodecagonal QC is stabilized by two roton instabilities, whose characteristic momenta $k_1$ and $k_2$ satisfy $k_2/k_1=\sqrt{2+\sqrt{3}}$~\cite{grossklags2025engineering,pupillo2020quantum}. As a result, apart from the zero-momentum contribution, the momentum distribution contains 24 dominant peaks, grouped into two sets of 12 peaks located on the roton rings with radii $k_1$ and $k_2$, respectively. 
In the numerical implementation, periodic boundary conditions are imposed and the system is discretized on a square grid. In a finite computational box, the momentum-space grid points generally do not coincide exactly with both roton rings simultaneously. Consequently, the QC peaks can only occupy the grid points closest to the ideal roton rings, as illustrated in Fig.~\ref{bijin}(a1). The resulting ratio between the actual peak momenta therefore deviates from the ideal value, leading to a noticeable distortion of the dodecagonal QC, as evidenced by the density profile shown in Fig.~\ref{bijin}(b1).

Nevertheless, the ideal quasicrystalline state can, in principle, be approached systematically by increasing the momentum resolution. Fig.~\ref{bijin} presents the converged numerical solutions obtained at different momentum resolutions. In an ideal dodecagonal QC, the angle between two neighboring momentum peaks on the same roton ring is $\pi/6$. In a finite computational domain, however, the discretization of momentum space limits the attainable angular resolution. For instance, the two adjacent peaks in Fig.~\ref{bijin}(a1) satisfy $\tan\alpha=3/5$, yielding a deviation from the ideal angle. As the momentum resolution is increased, $\alpha$ progressively approaches $\pi/6$, as shown in Fig.~\ref{bijin}(a1)$\sim$(a6). Correspondingly, the density profiles exhibit a clear convergence toward the ideal quasicrystalline state.  %The corresponding density distributions in spatial and momentum space are shown in Fig.~\ref{bijin} from left to right, from which it can be seen that our results are increasingly close to the ideal quasicrystalline states.

Table~\ref{table_energy} lists the superlattice period and energy obtained during the convergence toward the ideal quasicrystalline state. As $\alpha$ gradually approaches $\pi/6$, the spreads of the characteristic momenta $k_1$ and $k_2$ become progressively smaller, and the ratio between the average modulus value $\bar{k}_{1,2}$ of the twelve peaks along each momentum ring, i.e., $\bar{k}_2/\bar{k}_1$, gradually converges to $\sqrt{2+\sqrt{3}}$. Meanwhile, the superlattice period increases accompanied by a decrease in energy that exhibits clear convergence. These observations strongly indicate that the true ground state is realized in the infinite-period limit. In this limit, each characteristic momentum ring contains 12 equally spaced peaks, and the ratio between the two ring radii satisfies $k_2/k_1=\sqrt{2+\sqrt{3}}$. This configuration corresponds exactly to a dodecagonal quasicrystal, which is quasiperiodic rather than translationally periodic.

\begin{table}[!b]
	\renewcommand{\arraystretch}{1.5}
	\setlength{\tabcolsep}{11pt}
	\caption{The process of approximating a dodecagonal QC using square superlattices with increasing resolution in the momentum space. Here, the relevant parameters are $\rho_{2 \mathrm{D}}=50$, $a_s/a_{\mathrm{dd}}=0.75$, $\alpha=-0.00148$ and $\beta=14.69$ (cf. \textcolor[RGB]{0,120,0}{\large$\bullet$} in Fig.~2 in the main text). }
	\centering\scalebox{1}{
		\begin{tabular}{cccccccc}
			\toprule[0.5mm]
			
			$\tan\alpha$ & $\left|\alpha-\frac{\pi}{6}\right|/\frac{\pi}{6}$ & $\bar{k}_1$ & $\mathrm{Var}(k_1)$ & $\bar{k}_2$ & $\mathrm{Var}(k_2)$ & period & single-particle energy \\
			
			\midrule[0.2mm]
			
			2/3 & 12.3002\% & 0.3667 & 3.63$\times$10$^{-4}$ & 0.7075 & 1.02$\times$10$^{-4}$ & 64.03 & 0.95480 \\
			
			3/5 & 3.2125\% & 0.3656 & 2.67$\times$10$^{-5}$ & 0.7062 & 7.25$\times$10$^{-6}$ & 101.18 & 0.95256 \\
			
			4/7 & 0.8504\% & 0.3653 & 1.94$\times$10$^{-6}$ & 0.7057 & 5.18$\times$10$^{-7}$ & 138.32 & 0.95226 \\
			
			11/19 & 0.2286\% & 0.3653 & 1.39$\times$10$^{-7}$ & 0.7057 & 3.72$\times$10$^{-8}$ & 377.88 & 0.95225 \\
			
			15/26 & 0.0612\% & 0.3653 & 9.97$\times$10$^{-9}$ & 0.7058 & 2.67$\times$10$^{-9}$ & 516.14 & 0.95224 \\
			
			41/71 & 0.0164\% & 0.3653 & 7.16$\times$10$^{-10}$ & 0.7056 & 1.92$\times$10$^{-10}$ & 1411.18 & 0.95224 \\
			\bottomrule[0.5mm]
	\end{tabular}}
	\label{table_energy}
\end{table}

In addition to the approach described above, we have also employed the projection method~\cite{jiang2014numerical} to further verify that the dodecagonal quasicrystal is indeed the ground state. Within this framework, the calculations are performed in the 4D momentum space, where the QC is represented as a periodic structure. The resulting characteristic momenta and single-particle energies are consistent with those obtained using the previous method, providing independent confirmation of the dodecagonal quasicrystalline ground state. We also use this projection method to calculate the excitation spectrum of the QCs (see more details in Sec.~IV). Since it avoids the constraints imposed by periodic boundary conditions in the 2D space, the method enables us to successfully capture all Goldstone modes. It should be noted, however, that the projection method has certain limitations. First, because the basis vectors must be specified a priori, it is primarily applicable when the quasicrystalline nature of the ground state is already anticipated. Second, taking the dodecagonal QCs as an example, the calculations are carried out in 4D space, which significantly increases the computational cost and limits the accessible system size. Consequently, some high-energy and high-momentum excitations may not be determined with the same level of accuracy as the low-momentum and low-energy modes.

%In this method, our calculations are performed in 4D momentum space. We can obtain characteristic momenta and single-particle energy that consistent with those of the previous method, which further confirms that the ground state is dodecagonal QCs. We also employ this projection method when calculating the quasicrystal excitation spectrum. This method avoids the constraints imposed by periodic boundary conditions in 2D space, enabling us to successfully capture all Goldstone modes. 
%It should be noted that this method has inherent limitations. On the one hand, because it requires the prior selection of basis vectors, it can only be used to verify cases where it is relatively certain that the ground state is a quasicrystal. On the other hand, taking the dodecagonal QCs as an example, since the calculations are performed in 4D space, the size of computational space can not be increased significantly. For this reason, we may obtain some inaccurate excitations. 

\section{III. superfluidity of the dodecagonal quasicrystal}

In this section, we investigate the superfluid properties of the dodecagonal QCs. To characterize its superfluid response, we introduce the radial superfluid fraction defined as
\begin{equation}
f_{\mathrm{s}}(r_\perp)=\frac{n_{\mathrm{s}}(r_{\perp})}{n(r_{\perp})},
\label{eq:s10}
\end{equation}
with $n(r_{\perp})=\frac{1}{2\pi}\int_0^{2\pi}{\mathrm{d}\theta}\rho(\mathbf{r}_{\perp})$ being the average radial density, and $n_{\mathrm{s}}(r_{\perp})=\left(\frac{1}{2\pi}\int^{2\pi}_0\frac{\mathrm{d}\theta}{\rho(\mathbf{r}_{\perp})}\right)^{-1}$ the radial superfluid density~\cite{leggett1970can,leggett1998on}. 

\begin{figure}[!hb]
	\centering
	\includegraphics[width=0.6\columnwidth]{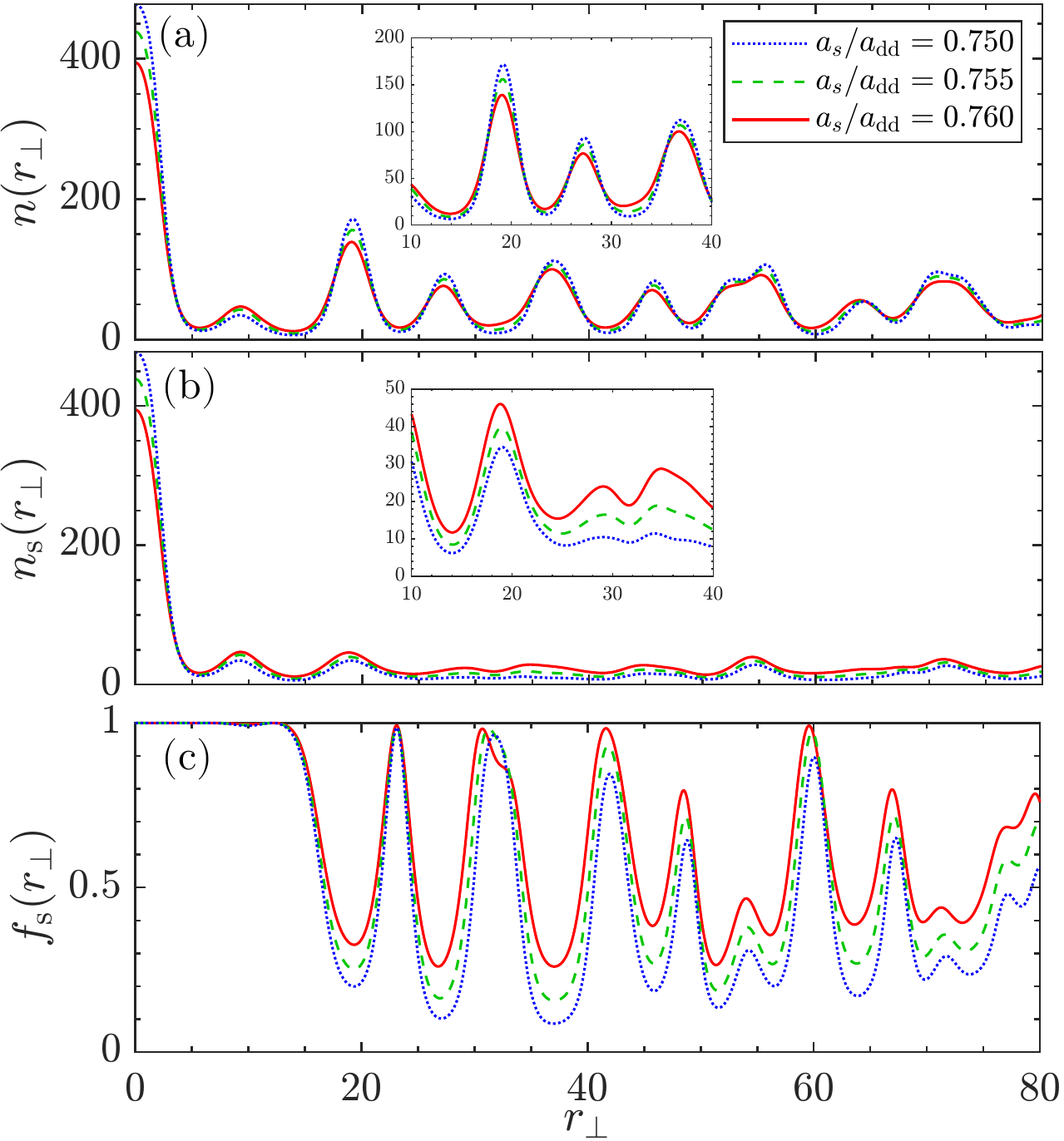}
	\caption{The spatial profiles of (a) the average radial density, (b) the radial superfluid density, and (c) the radial superfluid fraction for the dodecagonal QCs at $a_s/a_{\mathrm{dd}}=0.75$ (blue dotted line), $0.755$ (green dashed line), and $0.76$ (red line). The other parameters are $\rho_{2 \mathrm{D}}=50$, $\alpha=-0.00148$ and $\beta=14.69$. }
	\label{superfluid}
\end{figure}

As shown in Fig.~\ref{superfluid}, the contrast of $n(r_{\perp})$ decreases with increasing $a_s/a_{\rm dd}$, leading to an enhanced superfluid fraction. The finite superluid fraction demonstrates the coexistence of superfluidity and quasicrystalline orders in the dodecagonal QC phase, analogous to the coexistence of superfluid and crystalline orders in supersolids. This behavior is distinct from quasicrystals in alloys and other classical materials, where the superfluidity is absent. %To emphasize this distinction from their classical counterparts, we refer to the QCs predicted here as quantum QCs.

\section{IV. calculation method for excitations}

\begin{figure}[!ht]
	\centering
	\includegraphics[width=0.8\columnwidth]{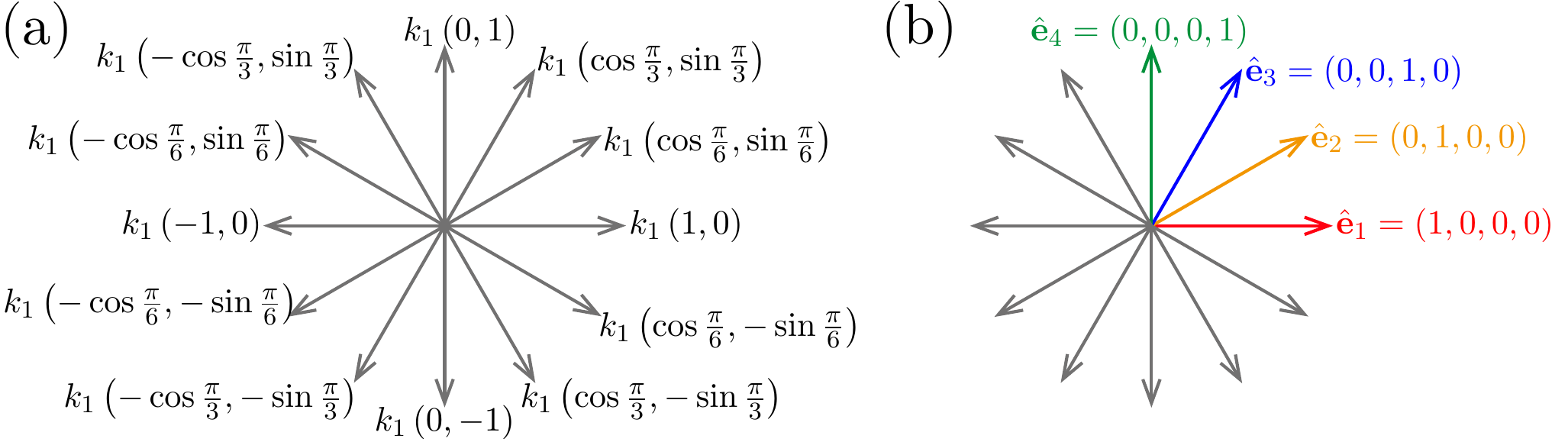}
	\caption{(a) The momenta of the twelve main peaks in the 2D momentum profile of the dodecagonal QCs (cf. bottom of the right column in Fig.~2 in the main text). Here $k_1$ denotes the modulus value of these momenta near the first roton minimum. (b) A corresponding set of the basis vectors in the 4D momentum space.}
	\label{2D_4D}
\end{figure}

In this section, we describe the numerical approach employed to calculate the excitation spectrum of dodecagonal QCs. As shown in Fig.~\ref{2D_4D}, the introduction of four basis vectors in the 4D space allows a one-to-one mapping between the 2D physical momentum space $\mathbf{k}_\perp=(k_x,k_y)^{\mathsf{T}}$ and the corresponding 4D reciprocal space $\mathbf{K}=(K_1,K_2,K_3,K_4)^{\mathsf{T}}$, i.e., $\mathbf{k}_\perp=\mathcal{S}\mathbf{K}$~\cite{jiang2014numerical}, where
\begin{equation}
	\mathcal{S}=
	k_1
	\begin{pmatrix} 1 & \cos\frac{\pi}{6} & \cos\frac{\pi}{3} & 0 \\
		 0 & \sin\frac{\pi}{6} & \sin\frac{\pi}{3} & 1 \end{pmatrix}
\end{equation}
is the transformation matrix with $k_1$ denoting the magnitude of the twelve principal momentum peaks in the momentum-space distribution of the dodecagonal QCs [cf. Fig.~2 in the main text as well as Fig.~\ref{fig:S5}(c)], and $K_{1,2,3,4}$ has been restricted to integer values. The excitation spectrum is then calculated in the corresponding 4D $K$-space.

The one-to-one mapping between the 2D physical space and the 4D reciprocal space allows the wave function $\psi(\mathbf{r})$ of the dodecagonal QCs to be expanded as
\begin{equation}
	\psi(\mathbf{r})=\sum_{\mathbf{K}}\tilde{\psi}(\mathbf{K})\mathrm{e}^{{i}\mathcal{S}\mathbf{K}\cdot\mathbf{r}}, 
\end{equation}
in terms of the integer-valued reciprocal-lattice indices $\mathbf{K}$, where $\tilde{\psi}(\mathbf{K})$ is the wave function in the 4D momentum space. Within the projection approach~\cite{jiang2014numerical}, it describes a periodic crystal in four dimensions, allowing the quasicrystal to be treated using methods developed for periodic systems. Substituting the above wave function into Eq.~(4) in the main text, the GPE can be reformulated in terms of $\tilde{\psi}(\mathbf{K})$ as follows,
\begin{equation}
	\begin{split}
		i\frac{\partial \tilde{\psi}(\mathbf{K})}{\partial t}
		=& \frac{1}{2}\left|\mathcal{S}\mathbf{K}\right|^2\tilde{\psi}(\mathbf{K})
		+g_{2\mathrm{D}}\sum_{\mathbf{K}_1+\mathbf{K}_2+\mathbf{K}_3=\mathbf{K}}
		\tilde{\psi}(\mathbf{K}_1)\tilde{\psi}(\mathbf{K}_2)\tilde{\psi}(\mathbf{K}_3) \\
		&+\gamma_{2\mathrm{D}}\sum_{\mathbf{K}_1+\mathbf{K}_2+\mathbf{K}_3+\mathbf{K}_4=\mathbf{K}}
		\tilde{\psi}(\mathbf{K}_1)\tilde{\psi}(\mathbf{K}_2)\tilde{\psi}(\mathbf{K}_3)\tilde{\psi}(\mathbf{K}_4)\\
		&+\sum_{\mathbf{K}_1+\mathbf{K}_2+\mathbf{K}_3=\mathbf{K}}
		\tilde{U}(\mathcal{S}\left(\mathbf{K}_1+\mathbf{K}_2\right))
		\tilde{\psi}(\mathbf{K}_1)\tilde{\psi}(\mathbf{K}_2)\tilde{\psi}(\mathbf{K}_3),
	\end{split}
    \label{GP_K}
\end{equation}
where $\tilde{U}(\mathbf{k}_\perp)=\tilde{U}_{\mathrm{DDI}}^{2\mathrm{D}}(\mathbf{k}_\perp)+\tilde{U}_{\mathrm{opt}}(\mathbf{k}_\perp)$.

For simplicity and without loss of generality, we take the ground-state wave function $\psi(\mathbf{r})$ of the 2D dodecagonal QCs to be real. The ground state $\tilde{\psi}(\mathbf{K})$ in the 4D $\mathbf{K}$-space is obtained by propagating Eq.~\eqref{GP_K} in imaginary time until convergence. Subsequently, the excitation spectrum is determined by solving the following Bogoliubov–de Gennes (BdG) equations,
\begin{equation}
	\begin{pmatrix} \mathcal{L}-\mu+\mathcal{X} & -\mathcal{X} \\
		\mathcal{X} & -(\mathcal{L}-\mu+\mathcal{X}) \end{pmatrix}
	\begin{pmatrix} \tilde{u}_{\nu\mathbf{q}}(\mathbf{K}) \\ \tilde{v}_{\nu\mathbf{q}}(\mathbf{K}) \end{pmatrix}
	=\omega_{\nu\mathbf{q}}
	\begin{pmatrix} \tilde{u}_{\nu\mathbf{q}}(\mathbf{K}) \\ \tilde{v}_{\nu\mathbf{q}}(\mathbf{K}) \end{pmatrix}, 
\end{equation}
where $\tilde{u}_{\nu\mathbf{q}}(\mathbf{K})$ and $\tilde{v}_{\nu\mathbf{q}}(\mathbf{K})$ are the Bogoliubov excitation amplitudes, $\omega_{\nu\mathbf{q}}$ donates the excitation frequency, and $\mu$ represents the chemical potential. The operators $\mathcal{L}$ and $\mathcal{X}$ are defined as
\begin{equation}
	\begin{split}
		\mathcal{L}\tilde{w}(\mathbf{K})
		=& \left(\frac{q^2}{2}+\mathbf{q}\cdot(\mathcal{S}\mathbf{K})+\frac{1}{2}|\mathcal{S}\mathbf{K}|^2\right)\tilde{w}(\mathbf{K})
		+g_{2\mathrm{D}}\sum_{\mathbf{K}_1+\mathbf{K}_2+\mathbf{K}_3=\mathbf{K}}
		\tilde{\psi}(\mathbf{K}_1)\tilde{\psi}(\mathbf{K}_2)\tilde{w}(\mathbf{K}_3) \\
		&+\gamma_{2\mathrm{D}}\sum_{\mathbf{K}_1+\mathbf{K}_2+\mathbf{K}_3+\mathbf{K}_4=\mathbf{K}}
		\tilde{\psi}(\mathbf{K}_1)\tilde{\psi}(\mathbf{K}_2)\tilde{\psi}(\mathbf{K}_3)\tilde{w}(\mathbf{K}_4)\\
		&+\sum_{\mathbf{K}_1+\mathbf{K}_2+\mathbf{K}_3=\mathbf{K}}
		\tilde{U}(\mathcal{S}\left(\mathbf{K}_1+\mathbf{K}_2\right))
		\tilde{\psi}(\mathbf{K}_1)\tilde{\psi}(\mathbf{K}_2)\tilde{w}(\mathbf{K}_3)
	\end{split}
\end{equation}
and
\begin{equation}
	\begin{split}
		\mathcal{X}\tilde{w}(\mathbf{K})
		=& g_{2\mathrm{D}}\sum_{\mathbf{K}_1+\mathbf{K}_2+\mathbf{K}_3=\mathbf{K}}
		\tilde{\psi}(\mathbf{K}_1)\tilde{\psi}(\mathbf{K}_2)\tilde{w}(\mathbf{K}_3)
		+\frac{3}{2}\gamma_{2\mathrm{D}}\sum_{\mathbf{K}_1+\mathbf{K}_2+\mathbf{K}_3+\mathbf{K}_4=\mathbf{K}}
		\tilde{\psi}(\mathbf{K}_1)\tilde{\psi}(\mathbf{K}_2)\tilde{\psi}(\mathbf{K}_3)\tilde{w}(\mathbf{K}_4)\\
		&+\sum_{\mathbf{K}_1+\mathbf{K}_2+\mathbf{K}_3=\mathbf{K}}
		\tilde{U}(\mathcal{S}\left(\mathbf{K}_1+\mathbf{K}_2\right)+\mathbf{q})
		\tilde{w}(\mathbf{K}_1)\tilde{\psi}(\mathbf{K}_2)\tilde{\psi}(\mathbf{K}_3),
	\end{split}
\end{equation}
respectively, where $\mathbf{q}=(q_x,q_y)$ represents the excitation momentum in the 2D physical space. Once the excitation spectrum is obtained, the dynamical structure factor can be straightforwardly evaluated according to
\begin{equation}
	S(\mathbf{q},\omega)=
	\sum_{\nu}\left|\sum_{\mathbf{K}}\left(\tilde{u}^*(\mathbf{K})-\tilde{v}^*(\mathbf{K})\right)\tilde{\psi}(\mathbf{K})\right|^2\delta(\omega-\omega_{\nu\mathbf{q}}).
    \label{eq:s17}
\end{equation}
The dynamical structure factor is a fundamental observable that characterizes the system's density fluctuations in response to the perturbation with a given momentum $\mathbf{q}$ and provides direct experimental access to the excitation spectrum~\cite{BECandSF,Exc2019Ferlaino,blakie2025dirac}. 

\begin{figure}[!ht]
	\centering
	\includegraphics[width=1\columnwidth]{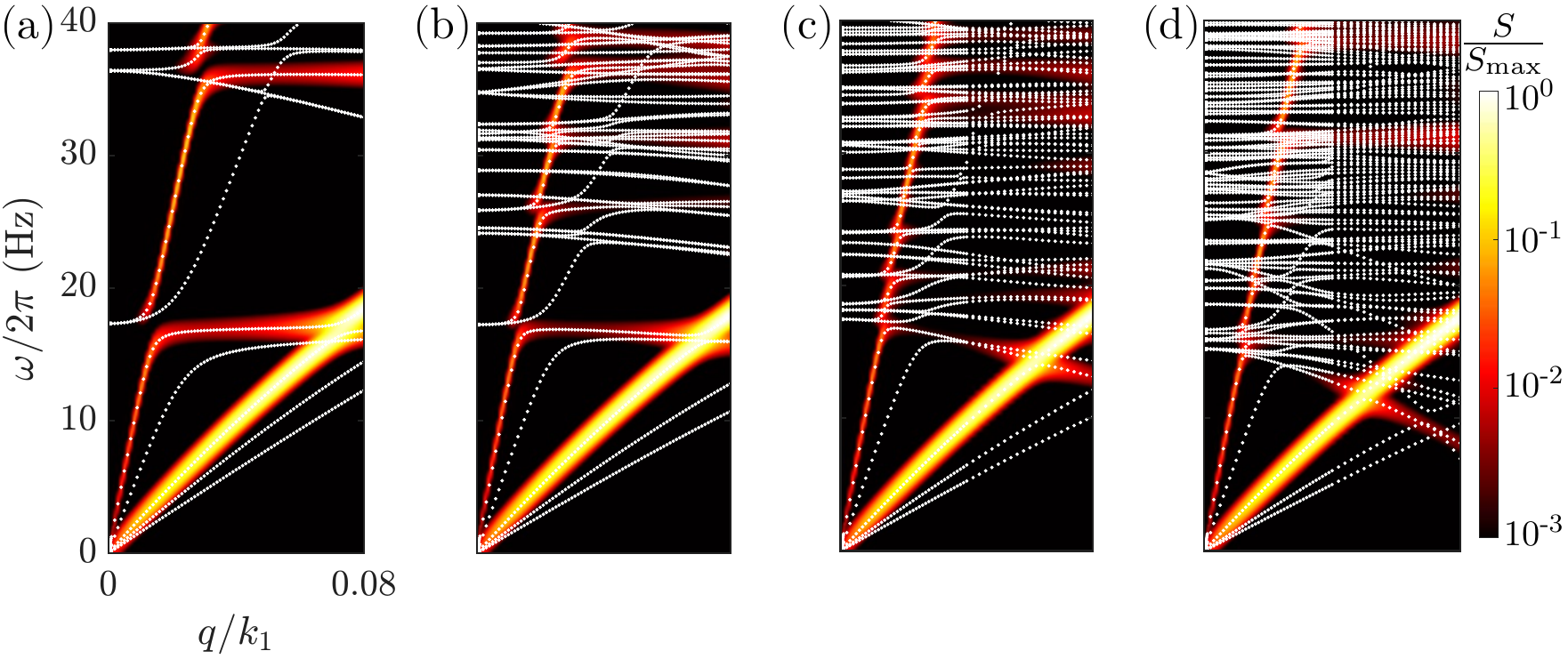}
	\caption{The excitation spectrum (white dotted lines) and dynamic structure factor (color intensity) of dodecagonal QCs obtained from calculations using different computational domains in the 4D momentum space, where the momenta have been limited to (a) $|K_{1,2,3,4}|\leqslant 3$, (b) $|K_{1,2,3,4}|\leqslant 5$, (c) $|K_{1,2,3,4}|\leqslant 6$, and (d) $|K_{1,2,3,4}|\leqslant 8$, respectively. Here, to enhance the visibility of the dynamic structure factor, the delta function $\delta(\omega-\omega_{\nu \mathbf{q}})$ in Eq.~(\ref{eq:s17}) has been broadened into a Gaussian with the frequency width $0.005\omega_z$. The excitation momentum $\mathbf{q}$ is taken along the $x$-axis in the 2D physical space. The ground-state quasicrystal parameters are identical to that marked by \textcolor[RGB]{0,120,0}{\large$\bullet$} in Fig.~2 in the main text.}
	\label{spectrum_3_5_6_8}
\end{figure}

Unlike periodic crystals, QCs lack translational symmetry and therefore do not possess well-defined Brillouin zones. Consequently, dodecagonal QCs exhibit dense distributions in both momentum space and their excitation spectra~\cite{DenseSpec1,DenseSpec2}. In practical calculations of the BdG equations, however, the 4D momentum $\mathbf{K}$ must be truncated at a finite cutoff due to computational limitations. This momentum cutoff introduces finite-size effects, causing the excitation spectrum to appear quasi-discrete rather than truly dense [see Fig.~\ref{spectrum_3_5_6_8}(a)]. As the momentum cutoff is increased, these finite-size effects are progressively reduced. As shown in Figs.~\ref{spectrum_3_5_6_8}(a)–(d), the dense nature of the excitation spectrum gradually emerges with increasing computational space. In particular, the five gapless modes converge rapidly and can therefore be determined accurately even with a relatively modest cutoff. Importantly, our approach captures not only the low-energy gapless branches but also high-frequency excitations. Furthermore, it can describe excitation properties beyond the linear regime at small $q$. In principle, the method is applicable to arbitrary momenta $q$, provided that a sufficiently large computational space is employed.

%Unlike regular crystals, the QCs do not have translational periodicity, leading to the absence of well-defined Brillouin zones. As a result, the dodecagonal QCs exhibit a dense distribution not only in the 2D momentum space but also in the excitation spectra~\cite{}. However, because of the limited computation resources in a realistic calculation of the BdG equation, one has to take a cutoff for the 4D momentum $\mathbf{K}$. Due to the finite-size effect caused by the momentum cutoff, the excitation spectrum presents a quasi-discrete characteristic instead of the ideal dense profile [see Fig.~\ref{spectrum_3_5_6_8}(a)]. However, this finite-size effect can be suppressed by increasing the momentum cutoff. As shown in Fig.~\ref{spectrum_3_5_6_8}(a$\sim$d), as the computation space goes up, the dense property of the excitation spectrum gradually emerges. In particular, the five gapless modes converge rapidly and thus can be addressed precisely even in a small computation space. It is worth highlighting that our approach can identify not only the lowest gapless branches but also the high-frequency excitations. Moreover, this approach is also capable describing the excitation behaviors beyond the linear regime of the gapless modes at small $q$. And, in principle, it also allows us to calculate the excitation spectra at an arbitrary $q$ with the requirement of sufficiently large computation space.

\section{V. Comparison between the locally gradual and globally sudden quench dynamics of the dBECs}
\begin{figure}[!hbt]
	\centering
	\includegraphics[width=\textwidth]{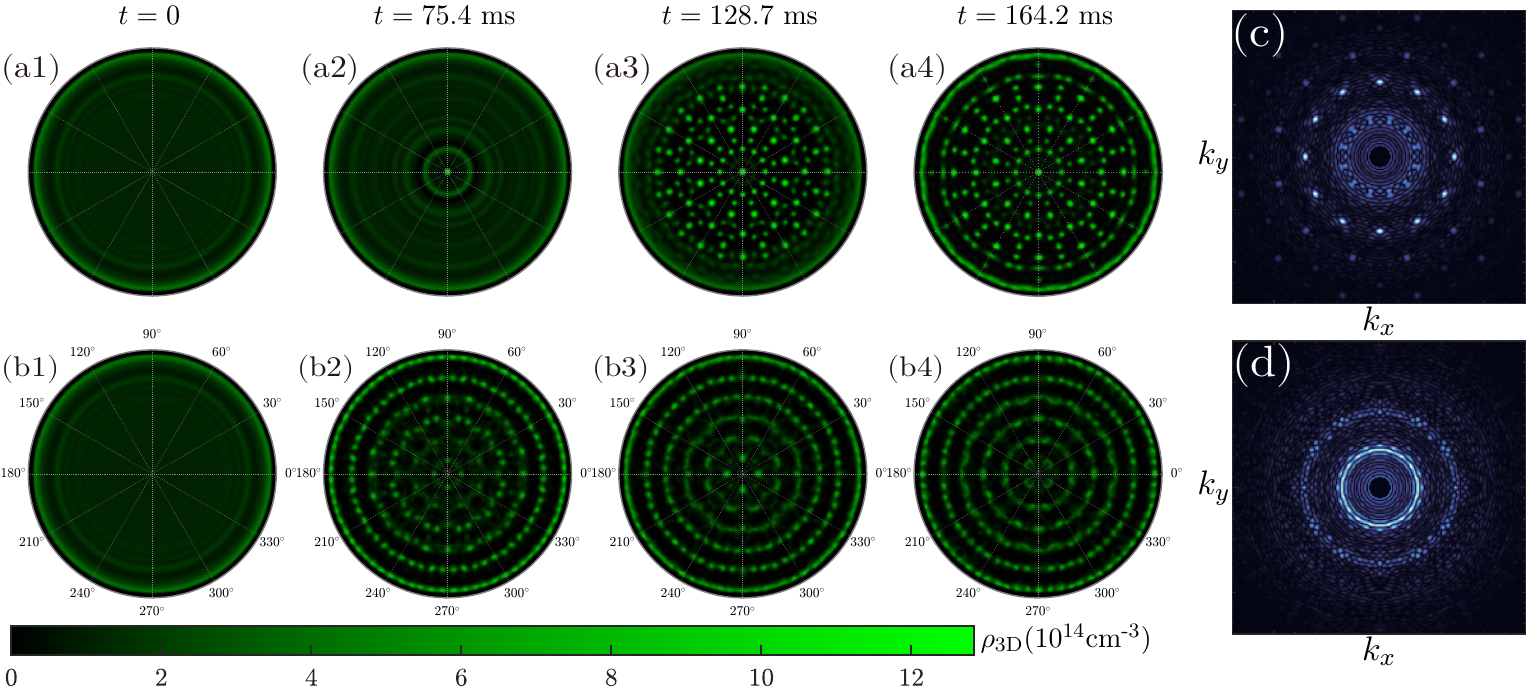}
	\caption{Dynamical evolution of the dBEC governed by the local and gradual quench with $v_0=0.4$ [(a1)-(a4), same as Fig.~4 in the main text] and by the global sudden quench [(b1)-(b4)]. Panels (c) and (d) show the corresponding momentum profiles of the states in panels (a4) and (b4), respectively, where the components near zero have been omitted for clarity of the peaks at finite momenta.}
	\label{fig:S5}
\end{figure}

In the main text we presented a local and gradual quench protocol to dynamically access a nearly-perfect QC. To further highlight its importance, we compare it to a global sudden quench of the interaction strength ${a_{s}}/{a_{\rm dd}}$ from 0.85 to 0.75 at the beginning. Fig.~\ref{fig:S5}(b1)-(b4) exhibits the corresponding dynamical evolution at the same times as Fig.~\ref{fig:S5}(a1)-(a4) (i.e., the same evolution process displayed in Fig.~4 in the main text), and Fig.~\ref{fig:S5}(d) shows the momentum profile of the state in Fig.~\ref{fig:S5}(b4). Notably, a global quench method is unable to produce QCs within a reasonable time scale as compared to our local and gradual quench approach. The reason could be that the entire condensate enters the QCs regime after a global quench, and then all the atoms tend to locally construct a discrete rotational symmetry. However, the competition between those local symmetry breakings hinders the building of an overall twelve-fold rotational symmetry. In contrast, the local and gradual quench offers a robust manner to deterministically establish the discrete rotational symmetry only at the center as a seed. Subsequently, the QCs gradually proliferate outwards until they fill the entire system. Considering that it remains challenging to produce a single large-size QC even in classical materials~\cite{ProduceQC1,ProduceQC2}, our approach presents a challenging yet possible viable route towards QCs.

\section{VI. experimental implementation of the proposed scheme}
In this section, we discuss a concrete experimental realization of our proposal by considering parameters appropriate for realistic atomic and optical fields. In the main text, we have set the two dimensionless parameters to $\alpha=-0.00148$ and $\beta=14.69$, which are defined as
\begin{equation}
	\alpha=\frac{2\tilde{\gamma}dm}{\hbar k_l}
	=-\frac{m\mu_+^2\mu_-^2\mathcal{E}_0^2}{4\hbar^4\epsilon_0 \lambda\Delta^2}
	=-0.00148
	\label{alpha}
\end{equation}
and
\begin{equation}
	\beta=\frac{d}{k_l \ell^2}
	=\frac{\lambda d}{288\pi^3 a_{\mathrm{dd}}^2}
	=14.69, 
	\label{beta}
\end{equation}
respectively. Here, for simplicity, we have assumed $\Delta_{+}=\Delta_{-}=\Delta$. Eq.~\eqref{alpha} and Eq.~\eqref{beta} define the driving field amplitude and the BEC-mirror distance via
\begin{equation}
	\mathcal{E}_0=\frac{2\hbar^2}{\mu_+\mu_-}\sqrt{\frac{0.00148\epsilon_0 \lambda}{m}}\Delta
\end{equation}
and
\begin{equation}
	d=\frac{4230.72\pi^3 a_{\mathrm{dd}}^2}{\lambda}. 
\end{equation}

\begin{comment}
In our schemes, large detuning condition and small phase condition should be satisfied~\cite{zhang2018long,zhang2021self,zhang2025nonreciprocal}, which leads to 
\begin{equation}
	\Omega_{\pm}=\frac{\mu_{\pm}\mathcal{E}_0}{\hbar}\ll\Delta
\end{equation}
and 
\begin{equation}
	\frac{\pi\mu_+^2}{\epsilon_0\hbar\lambda\Delta}n_{\mathrm{peak,2D}}\ll 1, 
\end{equation}
where $\Omega_{+}$ ($\Omega_{-}$) being the Rabi frequency of the forward (backward) light and $n_{\mathrm{peak,2D}}$ represents the 2D peak density during the dynamic process. 
Regarding the ``$\ll$" in the above equations, we can use a factor of 10 as a scale, and further obtain 
\begin{equation}
	\mu_{\pm}>20\hbar\sqrt{\frac{0.00148\epsilon_0 \lambda}{m}}
	\label{mu_lower}
\end{equation}
and
\begin{equation}
	\Delta>10\mu_+^2\frac{\pi n_{\mathrm{peak,2D}}}{\hbar\epsilon_0\lambda}, 
	\label{Delta_lower}
\end{equation}
Which provides lower bounds for the dipole matrix elements and the detunings. 
\end{comment}

\begin{figure}[!ht]
	\centering
	\includegraphics[width=0.8\columnwidth]{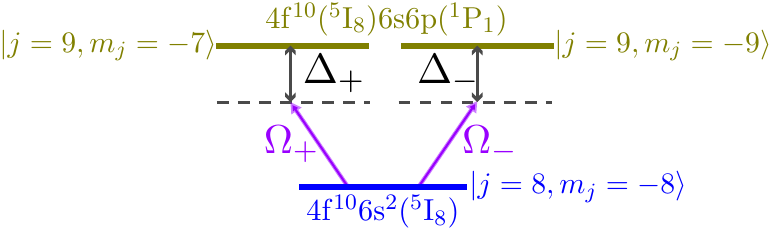}
	\caption{The V-type light-atom coupling configuration of $^{164}\text{Dy}$. Right (Left) circularly polarized light $\Omega_+$ ($\Omega_-$) couple the ground state $\ket{j=8,m_j=-8}$ with the excited state $\ket{j'=9,m_j'=-7}$ ($\ket{j'=9,m_j'=-9}$). %The electronic configuration of which are $4\mathrm{f}^{10}6\mathrm{s}^2(^{5}\mathrm{I}_8)$ and $4\mathrm{f}^{10}(^{5}\mathrm{I}_8)6\mathrm{s}6\mathrm{p}(^{1}\mathrm{P}_1)$ respectively. 
    }
	\label{level}
\end{figure}

To satisfy the above constraints, together with the large-detuning condition $\Omega_{\pm}=\frac{\mu_{\pm}\mathcal{E}_0}{\hbar}\ll\Delta$ and the small-phase condition $\frac{\pi\mu_+^2}{\epsilon_0\hbar\lambda\Delta}n_{\mathrm{peak,2D}}\ll 1$, we take $^{164}\text{Dy}$ BEC as an example. This system has a typical dipole length $a_{\mathrm{dd}}=131a_0$~\cite{bottcher2019dilute}, where $a_0$ is the Bohr radius. As illustrated in Fig.~\ref{level}, the atomic ground state is $\ket{j=8,m_j=-8}$ in the $4\mathrm{f}^{10}6\mathrm{s}^2(^{5}\mathrm{I}_8)$ manifold. The two excited states $\ket{j'=9, m_j'=-7}$ and $\ket{j'=9, m_j'=-9}$ are chosen from the $4\mathrm{f}^{10}(^{5}\mathrm{I}_8)6\mathrm{s}6\mathrm{p}(^{1}\mathrm{P}_1)$ manifold, allowing the transitions to be driven by an optical field with wavelength $\lambda\simeq 421\mathrm{nm}$. In this configuration, the relevant dipole matrix elements between the atomic ground and excited states are given by
%Taking $^{164}\text{Dy}$ BEC as an example, the typical dipole length of which is $a_{\mathrm{dd}}=131a_0$~\cite{bottcher2019dilute}, where $a_0$ is the Bohr radius. Thus we have $\omega_z=845~\mathrm{Hz}$, $\sigma_z=3.8~\upmu\mathrm{m}$ and $n_{\mathrm{peak,2D}}=6.49\times 10^{15}~\mathrm{m}^{\text{-}2}$. The electron configuration of ground state is $4\mathrm{f}^{10}6\mathrm{s}^2(^{5}\mathrm{I}_8)$, and we prepare it by an external magnetic field to $\ket{j=8,m_j=-8}$. Then as shown in Fig.~\ref{level}, light field $\Omega_{+}$ ($\Omega_{-}$) with wavelength $\lambda\simeq 421.172~\mathrm{nm}$ can be used to couple with the excited states $\ket{j'=9, m_j'=-7}$ ($\ket{j'=9, m_j'=-9}$) with electronic configuration being $4\mathrm{f}^{10}(^{5}\mathrm{I}_8)6\mathrm{s}6\mathrm{p}(^{1}\mathrm{P}_1)$, and the reduced dipole matrix element of the transitions is $\mu_{\mathrm{r}}=11.9 ea_0$~\cite{lu2011spectroscopy,chomaz2023dipolar,martin1978atomic,Kramida1999-gv}. The actual value of dipole matrix elements can be calculated via
\begin{equation}
		\mu_{\pm}=\begin{pmatrix} j' & 1 & j \\ -m_{j'} & m_{j'}-m_{j} & m_j \end{pmatrix}\mu_{\mathrm{r}}, 
\end{equation}
where $\left(\cdots\right)$ represents the Wigner 3-j symbol, and $\mu_{\mathrm{r}}=11.9 ea_0$ is the reduced dipole matrix element~\cite{lu2011spectroscopy,chomaz2023dipolar,martin1978atomic,Kramida1999-gv}. This yields $\mu_+=0.2207ea_0$ and $\mu_-=2.7300ea_0$. The constraints discussed are fulfilled for the following choice of parameters: $\omega_z=845\mathrm{Hz}$, $\rho_{\mathrm{peak,2D}}=6.49\times 10^{15}~\mathrm{m}^{\text{-}2}$, $\Delta=1.82\mathrm{GHz}$, $d=14.97~\upmu\mathrm{m}$ and a light intensity of $I_{0}=\frac{1}{2}c\epsilon_0 \mathcal{E}_{0,\mathrm{min}}^2=2.32\mathrm{mW}\cdot\mathrm{cm}^{\text{-}2}$. 

%It can be verified that these values satisfy Eq.~\eqref{mu_lower}. In this case, Eq.~\eqref{Delta_lower} gives the lower bound of $\Delta$ as $1.82~\mathrm{GHz}$. 
%When $\Delta$ takes the lower bound, the light intensity is 
%\begin{equation}
%	I_{0}=\frac{1}{2}c\epsilon_0 \mathcal{E}_{0,\mathrm{min}}^2
%	=\frac{0.296\pi^2 \hbar^2 c n_{\mathrm{peak,2D}}^2\mu_1^2}{m\lambda\mu_2^2}
%	=2.32~\mathrm{mW}\cdot\mathrm{cm}^{\text{-}2}. 
%\end{equation}
%Distance between dBEC and the mirror should be set to $d=14.97~\upmu\mathrm{m}$. 

In addition, we emphasize that $^{164}\text{Dy}$ represents just one possible candidate for experimental implementation. For example, $^{162}\text{Dy}$, $^{166}\text{Er}$ or other suitable atomic species can also be employed by properly adjusting the wavelength, light intensity, detunings, and other relevant physical parameters. %In addition, such a system can also be realized by $^{168}\mathrm{Er}$ or $^{166}\mathrm{Er}$. We need to select appropriate energy levels that satisfies Eq.~\eqref{mu_lower}, and then calculate the corresponding values for the other parameters.

\bibliography{mybib}